\documentclass[aps,pre,reprint,groupedaddress]{revtex4-2}
\usepackage{graphicx}
\usepackage{color}
\usepackage{amsmath}

\usepackage[T1]{fontenc}

\usepackage{ulem}

\begin{document}


\title{
Spatial quasiperiodic driving of a dissipative optical lattice and origin of directed Brillouin modes in a randomly diffusing cold atom cloud
}
\author{David Cubero}
\email[]{dcubero@us.es}
\affiliation{Departamento de F\'{\i}sica Aplicada I, Escuela Polit\'ecnica Superior, Universidad de Sevilla, Calle Virgen de \'Africa 7, 41011 Sevilla, Spain}
\author{Kefeng Jiang, Alexander Staron, Casey Scoggins, Daniel Wingert, Ian Dilyard, Stone Oliver, and Samir Bali}
\email[]{balis@miamioh.edu}
\affiliation{Department of Physics, Miami University, Oxford, Ohio 45056-1866,
USA}

\date{\today}
\begin{abstract}
Atoms confined in a three-dimensional dissipative optical lattice oscillate inside potential wells, occasionally hopping to adjacent wells, thereby diffusing in all directions. Illumination by a weak probe beam modulates the lattice, yielding propagating atomic density waves, referred to as Brillouin modes which travel perpendicular to the direction of travel of the probe. \textcolor{black}{The probe is made incident at a small angle relative to a lattice symmetry axis, yielding a driving potential perturbation whose spatial period is not a multiple of the period of the underlying optical potential, thus enabling exploration of the regime of space quasiperiodic drive.} 
\textcolor{black}{A new theory, based on the Fourier decomposition of the current into its atomic density wave contributions, reveals that unlike the previously studied time quasiperiodic case, wherein a lattice driven by two incommensurate frequencies may exhibit abrupt suppression in directed current as the driving transitions from quasiperiodic to periodic, a spatial-quasiperiodically driven lattice exhibits no such abrupt response.}
\textcolor{black}{Further, detailed modeling of spatial-quasiperiodically driven lattices reveals that directed propagation occurs} \textcolor{black}{not only as a consequence of velocity-matching between the propagating modulation and the average velocity of the atom oscillating inside a well as was previously reported in the literature, but,} \textcolor{black}{also as a distinct consequence of a new mechanism, namely,} \textcolor{black}{frequency-matching between the modulation frequency and the oscillation frequencies.} 
\textcolor{black}{A systematic measurement of the 
\textcolor{black}{transmitted probe} spectra as a function of off-axis probe angle is presented, which is consistent with the velocity-matching and frequency-matching predictions from the detailed model.} 
\end{abstract}

\maketitle

Light-induced forces on matter over wavelength and sub-wavelength spatial scales have wide applicability in quantum sensing and metrology~\cite{nat2021}, ranging from the novel design of periodic potential landscapes~\cite{porto2018,zoller2016,yavuz2013} to the innovative transport and sorting of particles~\cite{optica2020}. In particular, considerable interest has focused on the \textcolor{black}{Brownian} noise-induced directed motion of particles in the absence of a net force\textcolor{black}{~\cite{cubren16,hanggi2009,reimann2002}. These Brownian ratchets are central to any discussion of nanoscale transport, ranging from how natural biomolecular protein motors fuelled by stochastic collisions with surrounding water molecules transport intracellular cargo, to the design of artificial nanomachinery that can efficiently convert random environmental fluctuations to useful work~\cite{cubren16}. On a fundamental level the study of these Brownian ratchets enables an analysis of limitations imposed by the second law of thermodynamics in nonlinear stochastic processes~\cite{hanggi2009,reimann2002}.}
Cold atoms confined in dissipative optical lattices, where environmental noise in the form of spontaneous emission is significant, serve as ideally-controlled models that offer key insights into understanding Brownian ratchets~\cite{cubren16,cubero2012,gryn96,aspect1998,cecile99,grynreview,schiren02,renzoni02a,
renzoni03,grynpra2003,schiavoni2003,renzoniprl2004,gommers2005,gommers2006,
renzoni2007pra,renzoni2008prl,lebedev2009,cubero2010,cubero2011pre,umeapra12011,
umeapra22011,wichol12,samir22}. 

A dissipative optical lattice consists of counter-propagating laser beams tuned near atomic resonance that yield AC Stark-shifted ground state potential wells~\cite{metcalfbook}. Atoms in the lattice undergo the well-known process of Sisyphus cooling, and settle into these wells, where they oscillate with a vibrational frequency that is determined by the well-depth~\cite{grynreview}. The stochastic optical pumping processes associated with Sisyphus cooling also cause the atoms to occasionally transfer between adjacent wells, leading to spatial diffusion of the cold atom sample~\cite{grynreview,hodapp}. The introduction of a weak probe beam along a symmetry axis of the lattice, results in a time-periodic driving of the lattice that breaks the symmetry, causing directed atomic density waves to be set up. The directed propagation proceeds in the absence of a net force. These propagating atomic density waves are referred to as Brillouin modes in analogy to acoustic waves rippling through a fluid~\cite{gryn96,renzoni02a,grynpra2003,samir22}. 
\textcolor{black}{However, as noted earlier, whereas in fluids the particle interactions are sufficiently strong to support the acoustic wave propagation, in the dilute optical lattice the directed density wave propagation proceeds without involving any interaction between the atoms~\cite{gryn96}}. Recently, a noise-induced resonant enhancement of this directed propagation was observed~\cite{samir22}, and a new theory, based on the Fourier decomposition of the current into its atomic density wave contributions~\cite{cubero22}, was developed to explain this stochastic resonance.   This theory was also able to successfully predict the thresholds for the transition to the regime of infinite density in the cold atom setup \cite{lutren13,cubero22}.

\vspace{-1mm}
In this paper, we investigate the driving of the lattice by illuminating the cold atoms with  
a probe beam propagating at a slight angle to the lattice symmetry axis. In this case, the spatial period of the driving perturbation is now not an integer multiple of the period of the underlying lattice potential. This allows, at least theoretically, for the possibility of the two spatial driving frequencies to be in irrational ratio, which permits the exploration of spatial quasiperiodic driving in these systems, in analogy with the time quasiperiodic case, wherein the lattice is driven by two incommensurate frequencies \cite{cubren16,cubero2012,cubero2014,cubero2018,cub18}. In the present spatial quasiperiodic case, though, we will show that the generated directed current is not as sensitive to the nature of the driving as in the time quasiperiodic case, where true quasiperiodicity is able to suppress the directed motion. Here the transition from periodicity to quasiperiodicity is not observed to be sharp. However, the chosen setup, where the spatial periods of the underlying lattice and the driving are basically uncoupled, sheds light on how the Brillouin modes are generated. \textcolor{black}{Detailed modeling reveals that there are two distinct mechanisms responsible for the origin of directed propagation. The velocity-matching mechanism, where a current is generated when the propagating modulation coincides with the average velocity of the atom oscillating inside a well, was identified earlier~\cite{gryn96}, and in subsequent works~\cite{renzoni02a,grynpra2003,samir22}. Here, we identify a new mechanism, where a directed current is produced when there is a frequency-matching between the modulation frequency and the oscillation frequencies. We present a systematic measurement of the 
pump-probe 
spectra as a function of off-axis probe angle \textcolor{black}{(the lattice beams collectively serve as the pump)}, which corroborates the predictions from both mechanisms, without being able to rule out any one of them, or being able to resolve whether one is more dominant. Indeed, our investigation suggests that both mechanisms are equally at play.}   

 The paper is organized as follows. In Sec.~\ref{sec:models} we define the system model studied. New analytical results, based on a Fourier decomposition of the current, are discussed  in Sec.~\ref{sec:ana}. Numerical simulations and experiments are discussed in Sec.~\ref{sec:num} and \ref{sec:expt}, respectively. Finally, Sec.~\ref{sec:con} ends with the conclusions.  

\vspace{-6mm}
\section{System models}
\label{sec:models}
\begin{figure}[b]
\includegraphics[width=8cm]{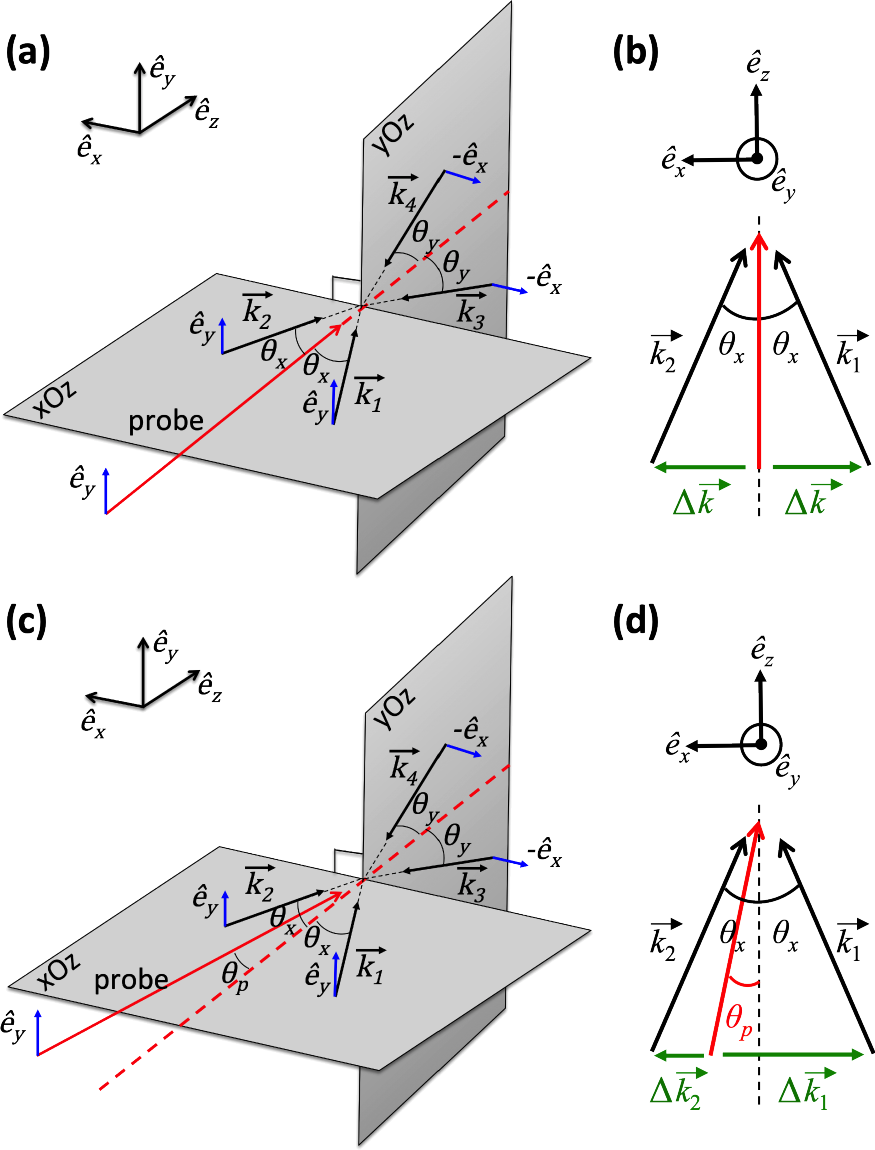}
\caption{
3D-lin$\perp$lin tetrahedral lattice illuminated by a weak probe. (a) and (b) depict the case of space periodic driving: The probe propagates along the $z$-axis,  which is the lattice symmetry axis. (c) and (d) show  the case where the space period of the driving is not a multiple of the period of the underlying  optical lattice: The probe propagates along a direction forming  an angle $\theta_p$ with the $z$-axis.
Here, $|\Delta{\vec{k}_1}| = k_0+k_l\sin\theta_p$ and $|\Delta{\vec{k}_2}| = k_0-k_l\sin\theta_p$, where  {$k_l$} is the laser wave number, and {$k_0=k_l\sin\theta_x$}. 
\label{fig:samir1}
}
\vspace{-3mm}
\end{figure}

\vspace{-2mm}
We consider atoms confined in a so-called 3D-lin$\perp$lin optical lattice \cite{grynreview}, formed by the superposition of four red-detuned laser beams $\vec{k}_{1-4}$, \textcolor{black}{of identical amplitudes $E_0$}, and frequency $\omega_l$ in a tetrahedral configuration, see Fig.~\ref{fig:samir1}(a).  
For $F_g=1/2 \rightarrow F_e=3/2$ atoms the lattice is formed by just two light-shifted ground state $\pm 1/2$-spin potentials\textcolor{black}{, denoted by $U_+$ and $U_-$}. An additional weak probe laser of \textcolor{black}{amplitude $E_p$ and} frequency $\omega_p$, forming an angle $\theta_p$ with the $z$-axis and with its polarization parallel to the $y$-axis, is added to drive the system out of equilibrium and put the atoms in directed motion \cite{gryn96}, $\theta_p = 0$ in Fig.~\ref{fig:samir1}(a, b) and $\theta_p \neq 0$ in Fig.~\ref{fig:samir1}(c, d).  In the experiments, this model is already a simplification, since the atoms have a more complex transition than $F_g=1/2 \rightarrow F_e=3/2$, but the theoretical results are still expected to provide good qualitative insight \cite{petsas99}.

Following previous studies \cite{gryn96,renzoni02b,renzoni03,renzoni02a,grynpra2003,samir22,cubero22}, we focus on movement along one of the directions, taken as the $x$-axis. The optical potential in each ground state, associated with the above setup is then given by (after taking $y=z=0$, \textcolor{black}{see appendix \ref{sec:apen0}})
\begin{eqnarray}
U_\pm(x,t)=\frac{U_0}{2}\Big[-\frac{3}{2}-\frac{1}{2}\cos(2 k_0 x) \pm  \cos(k_0 x)   \nonumber \\
+ \varepsilon_p\cos(\textcolor{black}{k_0} x+k_l\sin\theta_p  x -\delta_p t) \nonumber \\
+ \varepsilon_p\cos(\textcolor{black}{-k_0} x+k_l\sin\theta_p  x  -\delta_p t) \nonumber \\
\pm \varepsilon_p\cos(k_l\sin\theta_p  x  -\delta_p t)
\Big],
\label{eq:Upm:3D:0}
\end{eqnarray}   
where $k_0=k_l\sin\theta_x$, $k_l = \omega_l/c$ is the laser beam wave number, $\delta_p = \omega_p - \omega_l$ is a small probe detuning relative to the lattice ($\delta_p/\omega_l \ll 1$) \footnote{This definition of the probe detuning $\delta_p$ corrects a sign error in Ref.~\cite{samir22}.} \textcolor{black}{which modulates the lattice and is also referred to as the driving frequency},
$U_0=-16\hbar \Delta'_0/3 $,  
 with $\Delta'_0$ ($<0$) being the light-shift per lattice field,  and $ \varepsilon_p=E_p/(2E_0)$. 
The optical well-depth $U_0$ defines a vibrational frequency associated with an atom of mass $m_a$ oscillating at the bottom of a well,
\begin{equation}
\Omega_x=\textcolor{black}{k_0}\sqrt{3 U_0/2m_a} = 4\sin\theta_x\sqrt{|\Delta_0^\prime|\omega_r},
\label{eq:omx}
\end{equation}
where $\omega_r=\hbar k_l^2/(2m_a)$ is the recoil frequency.

The probe perturbation in (\ref{eq:Upm:3D:0}) appears in three terms: \textcolor{black}{For $\theta_p < \theta_x$, the first probe term in (\ref{eq:Upm:3D:0}) represents} a modulation propagating to the right with phase velocity $v_1=\delta_p/(k_0+k_l\sin\theta_p)$, \textcolor{black}{the second probe term a modulation propagating} to the left with velocity $v_2=-\delta_p/(k_0-k_l\sin\theta_p)$, and \textcolor{black}{the final probe term a modulation} moving with velocity $v_3=\delta_p/(k_l\sin\theta_p)$ \textcolor{black}{to the right or left depending on which side of the $z$-axis $\theta_p$ lies on; for $\theta_p$ as indicated in Fig.~\ref{fig:samir1}(c,d) $v_3$ is to the right}. Each of these terms will {excite atomic density waves. \textcolor{black}{The contributions of $v_1$ and $v_2$ to Brillouin propagation were discussed in \cite{gryn96,proc95}, but the role of $v_3$ in directed propagation has not been emphasized in the literature.} \textcolor{black}{The reason is that most of the previous works focused on the case of $\theta_p = 0$~\cite{renzoni02a,grynpra2003,samir22,cubero22}, which causes the last term in Eq.~\ref{eq:Upm:3D:0} for the potential to reduce to a pure function of time. In this case, no force is associated with the last term, and the term is neglected.} For the sake of simplicity \textcolor{black}{in keeping track of the contribution from $v_3$ versus the contributions from $v_1$ and $v_2$}, we consider them separately in the theoretical considerations that follow. 

Thus, the first system model to consider, denoted as case (a), is the following optical potential, which accounts for the third probe term in (\ref{eq:Upm:3D:0}) \textcolor{black}{moving with velocity $v_3$}, and is given by
\begin{eqnarray}
U_\pm^{(a)}(x,t)=\frac{U_0}{2}\Big[-\frac{3}{2}-\frac{1}{2}\cos(2 k_0 x) \pm  \cos(k_0 x)   \nonumber \\
\pm \varepsilon_p\cos(k_p x -\delta_p t+\phi_p)
\Big],
\label{eq:Upm:3D:a}
\end{eqnarray}   
where $k_p=k_l\sin\theta_p$ \textcolor{black}{is the driving wavenumber}, and $\phi_p$ is a probe phase which has been introduced for convenience in the analytical calculations to be presented below. 
Here the potential perturbation changes sign with the specific atomic state. Terms of this kind are also generated with a {$x$-polarized} probe.

\textcolor{black}{Next}, we consider the case of the following optical potential, denoted as case (b), which accounts for the first two probe terms in (\ref{eq:Upm:3D:0}), and is given by:
\begin{eqnarray}
U_\pm^{(b)}(x,t)=\frac{U_0}{2}\Big[-\frac{3}{2}-\frac{1}{2}\textcolor{black}{\cos}(2 k_0 x ) \pm  \cos(k_0 x )   \nonumber \\
+ \varepsilon_p\cos(k_p x - \delta_p  t+\phi_p)
\Big],
\label{eq:Upm:3D:b}
\end{eqnarray}   
\textcolor{black}{where, by} setting \textcolor{black}{the driving wavenumber $k_p$ to now be $k_p=|\Delta{\vec{k}_1}|=k_0+k_l\sin\theta_p$, or $k_p=|\Delta{\vec{k}_2}|=-(k_0-k_l\sin\theta_p)$},  
we can study the effects introduced by the first and second probe terms, respectively. 
See Figs.~\ref{fig:samir1} (b) and (d) for an illustration of $\Delta{\vec{k}_1}$ and $\Delta{\vec{k}_2}$. \textcolor{black}{Clearly,} if $\theta_p = 0$, $|\Delta{\vec{k}_{1,2}}| = |\Delta{\vec{k}}| = k_0$.

Note that the on-axis case $\theta_p=0$, recently investigated in Refs.~\cite{samir22,cubero22}, produces a space periodic perturbation, because the wave number of the driving field $k_p$ is the same as that of the underlying lattice, $k_0$. On the other hand, a probe angle $0<\theta_p<\pi/2$ such that $k_p/k_0$ is an irrational ratio produces a space quasiperiodic drive.

\textcolor{black}{We shall see below in Sec.~\ref{sec:num} that if we wish to determine whether the directed propagation changes abruptly as we transition between space periodic and quasiperiodic driving, as was observed in the previously studied time quasiperiodic case for a lattice driven by two incommensurate frequencies~\cite{cubren16,cubero2012,cubero2014,cubero2018,cub18}, it is advantageous to focus on system model (a). On the other hand, if the focus is on trying to understand how modes of directed propagation are generated (as was the case in previous works, e.g. Ref.~\cite{gryn96,proc95}), we may choose to study either system model (a) or (b) because they both yield similar results - in fact, we shall see in Sec.~\ref{sec:num} that system model (b) is mathematically simpler to analyze, hence the more favorable choice. This is probably another reason why the role of the $v_3$-term and system model (a) has been historically underemphasized - the above mentioned previous works were focused on Brillouin transport, not on the transition to quasiperiodicity where the $v_3$-term deserves most attention.} 

In the semi-classical approximation \cite{petsas99},  the atoms in the ground state  $|\pm\rangle$ are described by the phase space density $P_{\pm}(x,p,t)$  at the position $x$ with momentum $p$, which satisfies the following coupled Fokker-Planck equations 
\begin{eqnarray}
\left[ \frac{\partial}{\partial t} + \frac{p}{m_a}\frac{\partial}{\partial x} 
-U_{\pm}^\prime(x)\frac{\partial}{\partial p}\right] P_{\pm}= \nonumber\\
-\gamma_{\pm}(x) P_{\pm}+\gamma_{\mp}(x) P_{\mp} 
+\frac{\partial^2}{\partial p^2}\left[D_0 P_{\pm}\right],
\label{eq:fp}
\end{eqnarray}
where $U_{\pm}^\prime=\partial U_{\pm}/\partial x$,  and
\begin{equation}
\gamma_{\pm}(x)=g_0\pm g_1\cos(k_0 x) + g_2\cos(2 k_0 x) 
\label{eq:gammapm}
%
\end{equation}
are the transition rates between the ground state sublevels, defined in terms of $\Gamma'$, the photon scattering rate per lattice beam, as   $g_0=2\Gamma'/3$, $g_1=8\Gamma'/9$, $g_2=2\Gamma'/9$,
 and $D_0=5\hbar^2k_0^2\Gamma'/18$ 
 is a noise strength describing the random momentum jumps that result from the interaction with the photons. As in \cite{cubero22}, we are neglecting the probe contribution to the transition rates, the radiation forces, and noise terms, since their effect is observed to be small in the simulations. \textcolor{black}{Note that this approach is in contrast to the qualitative arguments presented in \cite{proc95} where the Brillouin-like directed motion was attributed to a synchronization of the 
probe-induced potential well depth modulation with the probe-modified transition rates between the ground state sub-levels of the $F_g = 1/2 \rightarrow F_e  = 3/2$ atom.
  }

Directed motion is characterized by the current, defined as the average atomic velocity
\begin{eqnarray}
&\langle v \rangle = \lim_{t\rightarrow\infty} \frac{\langle [x(t)-x(0)] \rangle }{t} = \nonumber\\
 &\lim_{t\rightarrow\infty}  \frac{1}{t} \int_0^t \!\! dt' \int \!\! dx \int \!\! dp \, \frac{p}{m_a}[P_+(x,p,t')+P_-(x,p,t')].
\label{eq:currdef}
\end{eqnarray}

\textcolor{black}{In Sec.~\ref{sec:ana} below, we present analytical expressions for the current which enable us to quantify the individual contributions from various atomic density modes excited by the probe.
These analytical expressions help us understand the plots generated by numerical solution of equation~\ref{eq:fp}, which are presented in Sec.~\ref{sec:num}.}   

\section{Analytical results}
\label{sec:ana}
Following the method presented in Ref.~\cite{cubero22}, we Fourier-decompose the current in (\ref{eq:currdef}) into contributions arising from atomic density modes excited by the probe. Using $l$, $n$, $m$ to denote the mode numbers, so that the mode has a frequency $\omega=l\delta_p$ and wave number $k=nk_0+mk_p$, we obtain,  
in terms of the amplitudes of the excited atomic density waves,
\begin{eqnarray}
P^{\pm}[l,n,m]=\frac{\delta_p}{2\pi}\int_0^{2\pi/\delta_p}\!\!\! dt \, e^{-i l \delta_p t}\int \!dx \,  e^{i (n k_0+m k_p) x} \nonumber \\
\int \!dp \,  P_{\pm}(x,p,t),
\label{eq:pfourier:def}
\end{eqnarray}
\textcolor{black}{expansions for the current. }

\textcolor{black}{The full expressions for the currents $\langle v \rangle_\mathrm{(a)}$ and $\langle v \rangle_\mathrm{(b)}$ for the setups defined by (\ref{eq:Upm:3D:a}) and (\ref{eq:Upm:3D:b}) for the $U^{(a)}$ and $U^{(b)}$ systems, respectively, are given in appendix \ref{sec:apen}.}
\textcolor{black}{Interestingly, the expression for the special case $k_p = k_0$ {\color{black} in system  $U^{(a)}$}  has extra terms, highlighted in blue (light grey) in (\ref{eq:3Dlin:a:per}), that cannot be obtained from the  {\color{black} more general quasiperiodic case ($k_p/k_0$ irrational)} in equation (\ref{eq:3Dlin:a:quasi}) 
just by simply taking the limit $k_p\rightarrow k_0$}. 

\textcolor{black}{Before commenting further on Eqs.~(\ref{eq:3Dlin:a:quasi})--(\ref{eq:3Dlin:a:per}) for the $U^{(a)}$ system, 
it is instructive to consider the analytic mode expansion for, say, the case of  $U^{(b)}$, which has been qualitatively discussed previously in Refs.~\cite{gryn96,proc95}: Below
we restrict ourselves to those terms that are {\cal O}($\epsilon_p$).}
 \begin{eqnarray}
{\langle v \rangle_\mathrm{(b)}} =  \langle v \rangle^{(0)}+ \frac{m_a}{m_a F_0 g_1 - 2 D_0 k_0}   \nonumber \\
 \times  \Bigg[\frac{\text{Im}\left[e^{i\phi_p}P_+[1,-2,1]\right] F_0 F_p (2k_0-k_p) }{2 m_a k_p}  \nonumber \\
+\frac{\text{Im}\left[e^{i\phi_p}P_+[1,-1,1]\right] F_0 F_p (-k_0+k_p) }{ m_a k_p}  \nonumber \\ 
+\frac{ \text{Im}\left[e^{i\phi_p}P_+[1,0,1]\right] 2 F_p k_0 \delta_p^2}{k_p^2}  \nonumber \\
+\frac{\text{Im}\left[e^{i\phi_p}P_+[1,1,1]\right] F_0 F_p (k_0+k_p) }{m_a k_p}  \nonumber \\
 -\frac{\text{Im}\left[e^{i\phi_p}P_+[1,2,1]\right]  F_0 F_p (2k_0+k_p) }{2 m_a k_p} \Bigg],
 \label{eq:3Dlin:b}
\end{eqnarray}
{\color{black}where $\langle v \rangle^{(0)}$, given in Eq.~(\ref{eq:v0}), contains intrinsic terms not related to the probe}, and the force amplitudes are given by $F_0=k_0U_0/2$ and $F_p=U_0\varepsilon_p k_p/2$.
Equation (\ref{eq:3Dlin:b}) is valid for an arbitrary $k_p$, including the periodic case $k_p=k_0$, which was validated numerically in Ref.~\cite{cubero22}. 
\textcolor{black}{
The expansion (\ref{eq:3Dlin:b}) provides a precise analytical decomposition into the current's contribution of each excited atomic wave. Despite its apparent complexity, when combined with a numerical calculation of the atomic wave amplitude, it is of great use to highlight the behavior of an individual atomic mode when a parameter is varied in the system, thus providing valuable physical insight, as it will be shown in the following section. }%

\textcolor{black}{
From simple inspection, one can deduce from (\ref{eq:3Dlin:b}) that the current diverges when the common denominator, $m_a F_0 g_1 - 2 D_0 k_0$, vanishes, a phenomenon known as going into the regime of infinite density \cite{lutren13,cubero22}. 
On the other hand, the current contributions from all modes except $[1,0,1]$ are seen from (\ref{eq:3Dlin:b})   to vanish for specific values of the probe wavelength $k_p$ (the contribution of $[1,-2,1]$ at $k_p=2k_0$ and so on). 
}\textcolor{black}{
Finally, we shall see in the following section, in Fig.~\ref{fig:sims5}, that the $[1,0,1]$ term in (\ref{eq:3Dlin:b}), traveling at the same speed as the probe perturbation, is the dominant Brillouin mode. } 

\textcolor{black}{We may now compare the expansion in (\ref{eq:3Dlin:b}) for case (b) with the expansions in Eqs.~(\ref{eq:3Dlin:a:quasi})--(\ref{eq:3Dlin:a:per}) for case (a)}. We  \textcolor{black}{conclude that}
the main difference between the cases (a) and (b), defined by (\ref{eq:Upm:3D:a}) and  (\ref{eq:Upm:3D:b}), respectively,  is that case (a)  provides different expressions for the generic case $k_p/k_0$ irrational and the special case $k_p=k_0$: In the expansion for the periodic case $k_p=k_0$, given in (\ref{eq:3Dlin:a:per}), there are more terms, highlighted in blue (light grey) in (\ref{eq:3Dlin:a:per}) ---thus more atomic wave modes activated--- than in the quasiperiodic case where  $k_p/k_0$ is an irrational ratio, (\ref{eq:3Dlin:a:quasi}). 

Furthermore, (\ref{eq:3Dlin:a:quasi}) shows some apparent singularities in the form of coefficients with denominators proportional to $k_0-k_p$ or $k_0-2k_p$, and thus apparently problematic in the periodic limits $k_p\rightarrow k_0$ and $k_p\rightarrow 2k_0$. \textcolor{black}{Importantly,} this could suggest a special sensitivity to the periodic/quasiperiodic transition, similar to that observed in the case of time quasiperiodicity \cite{cubren16,cubero2012,cubero2013,cubero2014,cubero2018,cub18}, where the large sensitivity in the system response to time quasiperiodic forces is known to yield sub-Fourier resonances.
\textcolor{black}{However, our numerical results, reported in the following section, shows that this is not the case, and the transition is smooth. 
 The analytical expansions, though, still provide a useful decomposition into atomic waves of directed transport, and are used  in the following sections to interpret the atomic transport provoked by the probe.}

\section{Numerical results}
\label{sec:num}

Numerical solutions of the equation (\ref{eq:fp}) are obtained by generating a large number of individual atomic trajectories $x_j(\sigma_j(t),t)$, where $\sigma_j(t)=+1$ or $-1$ is the occupied state at time $t$ in that trajectory, using a stochastic algorithm \cite{kloeden}. Averages are computed using over $1.5\times 10^6$ trajectories. Following Ref.~\cite{cubero22}, the atomic mode amplitudes (\ref{eq:pfourier:def}) are calculated via the formula,
\begin{eqnarray}
&P_\pm[l,n,m]= \lim_{l'\rightarrow\infty} \frac{\delta_p}{2\pi l' N} \times\nonumber\\
&\sum_{j=1}^{N}\int_0^{2\pi l'/\delta_p}\!\!\! dt \,  e^{i [(n k_0+m k_p) x_j(\sigma_j(t),t) - l \delta_p t] }\delta_{\sigma_j(t),\pm1}.
\label{eq:pfourier:def3}
\end{eqnarray}

In all simulations, units are defined such that $m_a=\hbar=k_l=1$, {\color{black} so that the atomic recoil velocity $v_r=\hbar k_l/m_a=1$ and recoil frequency $\omega_r=\hbar k_l^2/2m_a=1/2$}. In these units, the optical lattice parameters were fixed to $U_0=200$, $\Gamma'=2.85$. 

We start by first studying the system sensitivity to the transition between space periodic and quasiperiodic driving. We choose the $U^{(a)}$ system in (\ref{eq:Upm:3D:a}) {(i.e., case (a))},  because it gives different expansions in the periodic $k_p=k_0$ and quasiperiodic  ($k_p$ and $k_0$ incommensurate) cases. Specifically, (\ref{eq:3Dlin:a:quasi}) indicates a potential singularity when $k_p\rightarrow k_0$ in the  mode $\mathrm{Re}[P_+[1, -1, 1]]$ due to the presence of $(k_0-k_p)$ in the denominator of the coefficient \textcolor{black}{(this term is highlighted in blue (light grey) in (\ref{eq:3Dlin:a:quasi})). Potential singularities may also surface when $k_p \rightarrow -k_0, \pm 2k_0$ due to ($k_0 + k_p$) and ($2k_0 \pm k_p$) occurring in the denominator of several terms in (\ref{eq:3Dlin:a:quasi}), as mentioned above.}

Figure~\ref{fig:sims1} shows the current and the mode contributions to the current, given by (\ref{eq:3Dlin:a:quasi}), as a function of $k_p$ 
\textcolor{black}{in the vicinity of $k_0$}. 
\begin{figure}[h]
\includegraphics[width=8cm]{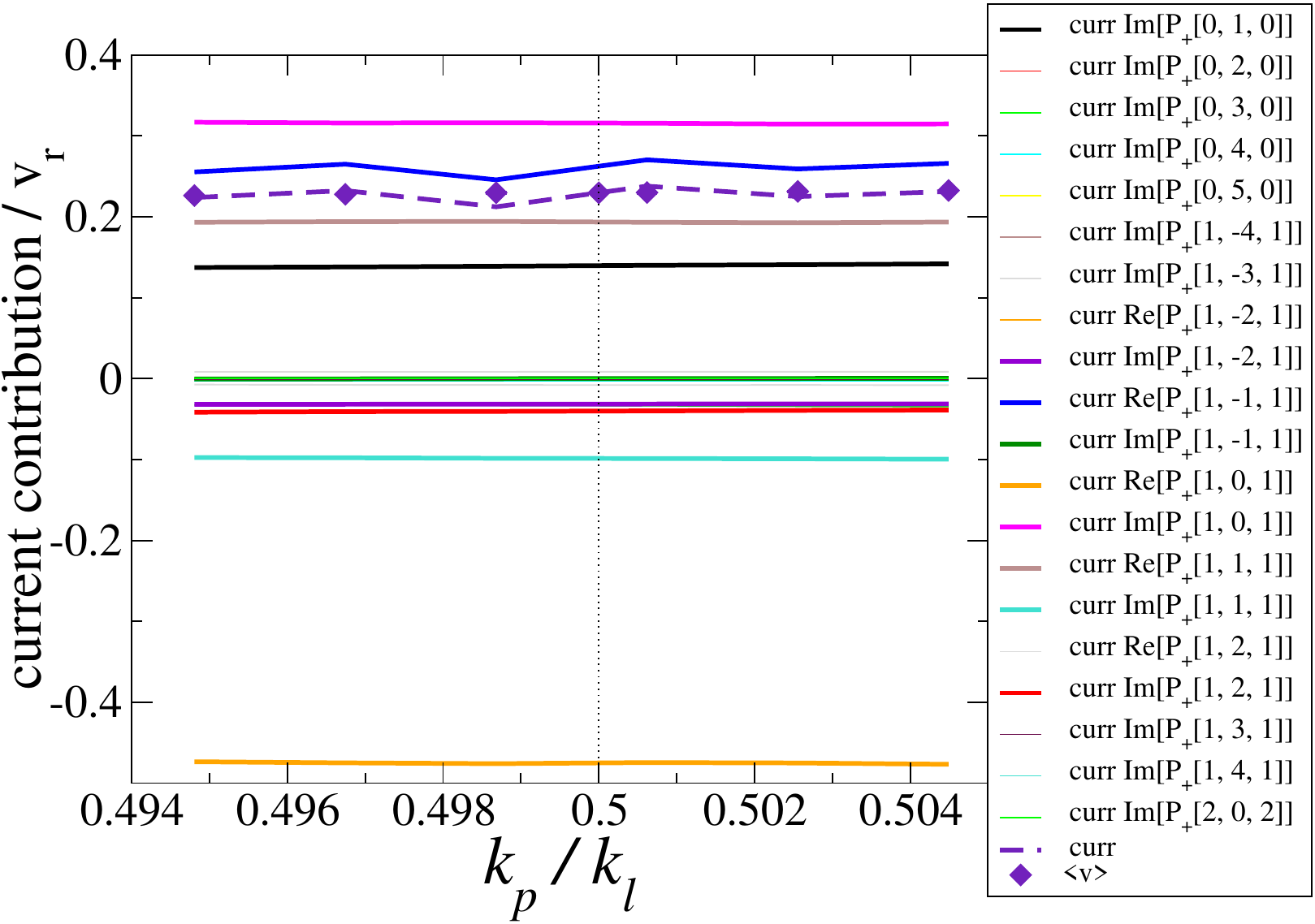}
\caption{
Smooth space quasiperiodicity. Current and mode contributions to the current as a function of the driving wavenumber $k_p$\textcolor{black}{$=k_l\sin\theta_p$} {for the $U^{(a)}$ system in (\ref{eq:Upm:3D:a}) (i.e., case (a))}, with $\delta_p/(2\omega_r)=7.5$, $\epsilon_p=0.8$, $\theta_x=30^0$ (thus $k_0=0.5 k_l$\textcolor{black}{, shown by the dotted vertical line}), and $\phi_p=0$. Each mode $(l,n,m)$ has a frequency $\omega=l\delta_p$ and wave number $k=nk_0+mk_p$.  Mode amplitudes are measured in the simulation via (\ref{eq:pfourier:def3}), and their precise contribution to the current determined using (\ref{eq:3Dlin:a:quasi}). \textcolor{black}{All the modes in (\ref{eq:3Dlin:a:quasi}) must be plotted to be sure of the smoothness of the transition between space periodicity and quasiperiodicity.} \textcolor{black}{The dashed line is the sum of all current contributions, and the diamonds are the current calculated from its definition (\ref{eq:currdef}).}
\label{fig:sims1}
}
\end{figure}
No mode or contribution in Fig.~\ref{fig:sims1}} is  observed to act abruptly, on the contrary, they all are seen to behave smoothly. The potential singularity in the mode $\mathrm{Re}[P_+[1, -1, 1]]$ (the blue/light-grey highlighted term in (\ref{eq:3Dlin:a:quasi})) does not actually occur, because the mode amplitude itself tends to zero in the limit $k_p\rightarrow k_0$, such that its current contribution is finite, \textcolor{black}{as illustrated by} Fig.~\ref{fig:sims1}.

Similar features are observed near $k_p=2k_0$, demonstrating that the transition from space quasiperiodicity to periodicity is smooth.

 %



Next, in order to understand how Brillouin \textcolor{black}{resonant} modes are generated, we study the atomic waves when varying the driving frequency $\delta_p$. {\color{black} We may study either one of the $U^{(a)}$ or $U^{(b)}$ systems because they both yield similar behavior. Here, we choose system (b)
because it has a simpler expression for the current as compared to system (a): Eqn. (\ref{eq:3Dlin:b}) is valid for arbitrary $k_p$ and has fewer terms than the expressions (\ref{eq:3Dlin:a:quasi}) or (\ref{eq:3Dlin:a:per}). The $U^{(b)}$ system yields a} series of resonances, that is, local maxima at certain values of the driving frequencies, 
allowing for a proper rationalization of the transport mechanisms and the experimental results. 

\begin{figure}[h]
\includegraphics[width=8cm]{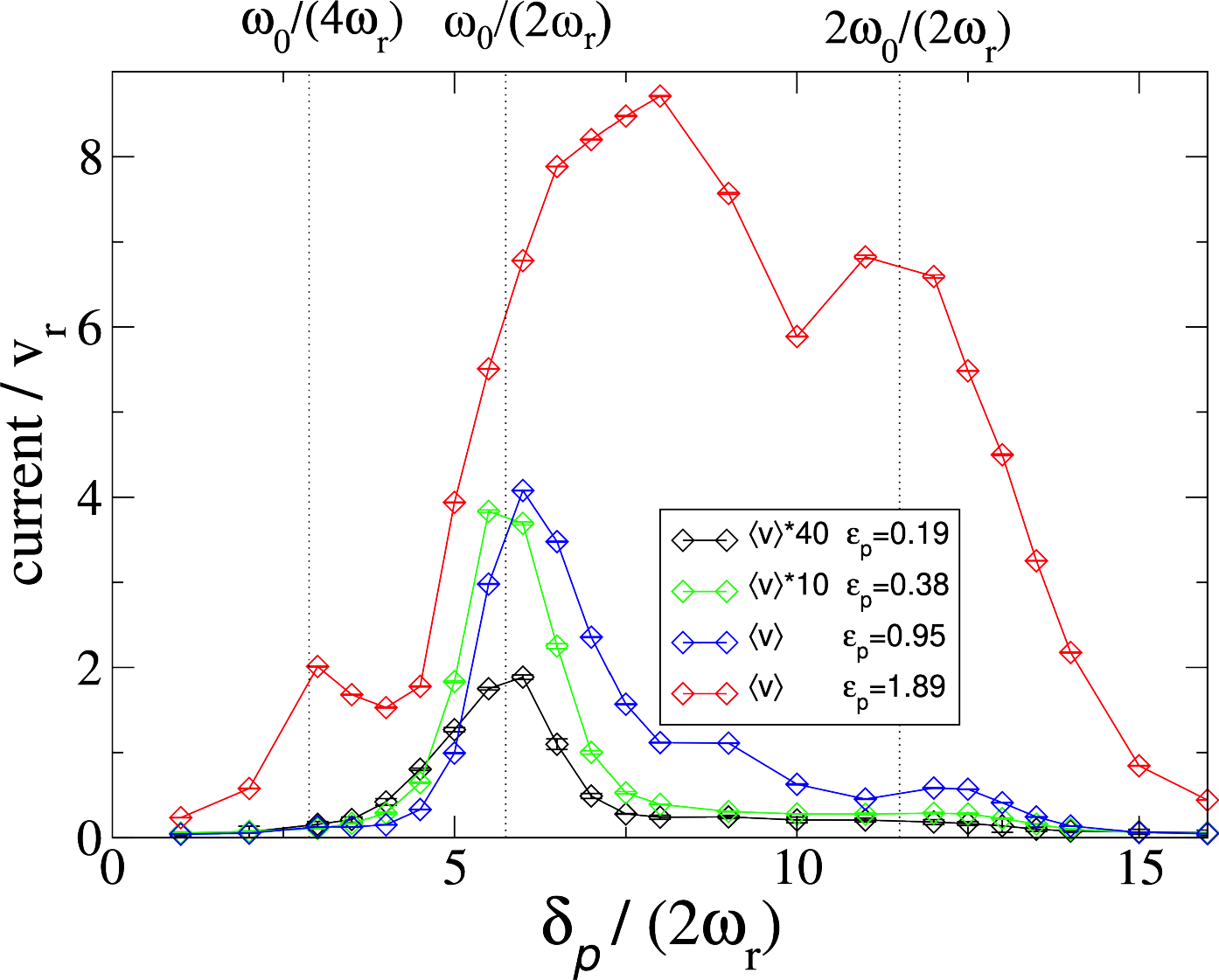}
\caption{
Space periodic driving of the $U^{(b)}$ system in (\ref{eq:Upm:3D:b}) (i.e., case (b)): The atomic current (open diamonds) is plotted for the case of space periodic driving as a function of the driving frequency $\delta_p=\omega_p - \omega_l$, in the $U^{(b)}$ system defined by (\ref{eq:Upm:3D:b}), with $\theta_x=25^0$, $\phi_p=\pi$, for several values of the driving amplitude $\varepsilon_p$. The solid lines are merely to guide the eye. The driving wave number is fixed to $k_p=k_0=k_l \, \mbox{sin}\,\theta_x$. Dotted vertical lines 
are placed at $\omega_0/(2\omega_r)=5.75$, $\omega_0/2$, and $2\omega_0$, where $\omega_0$ is identified with the intrinsic vibrational frequency $\Omega_x$ of the lattice (see text). We see that resonances at multiples and sub-multiples of the intrinsic frequency $\omega_0$ are 
predicted.
\label{fig:sims3}
}
\end{figure}

\begin{figure}[h]
\includegraphics[width=8cm]{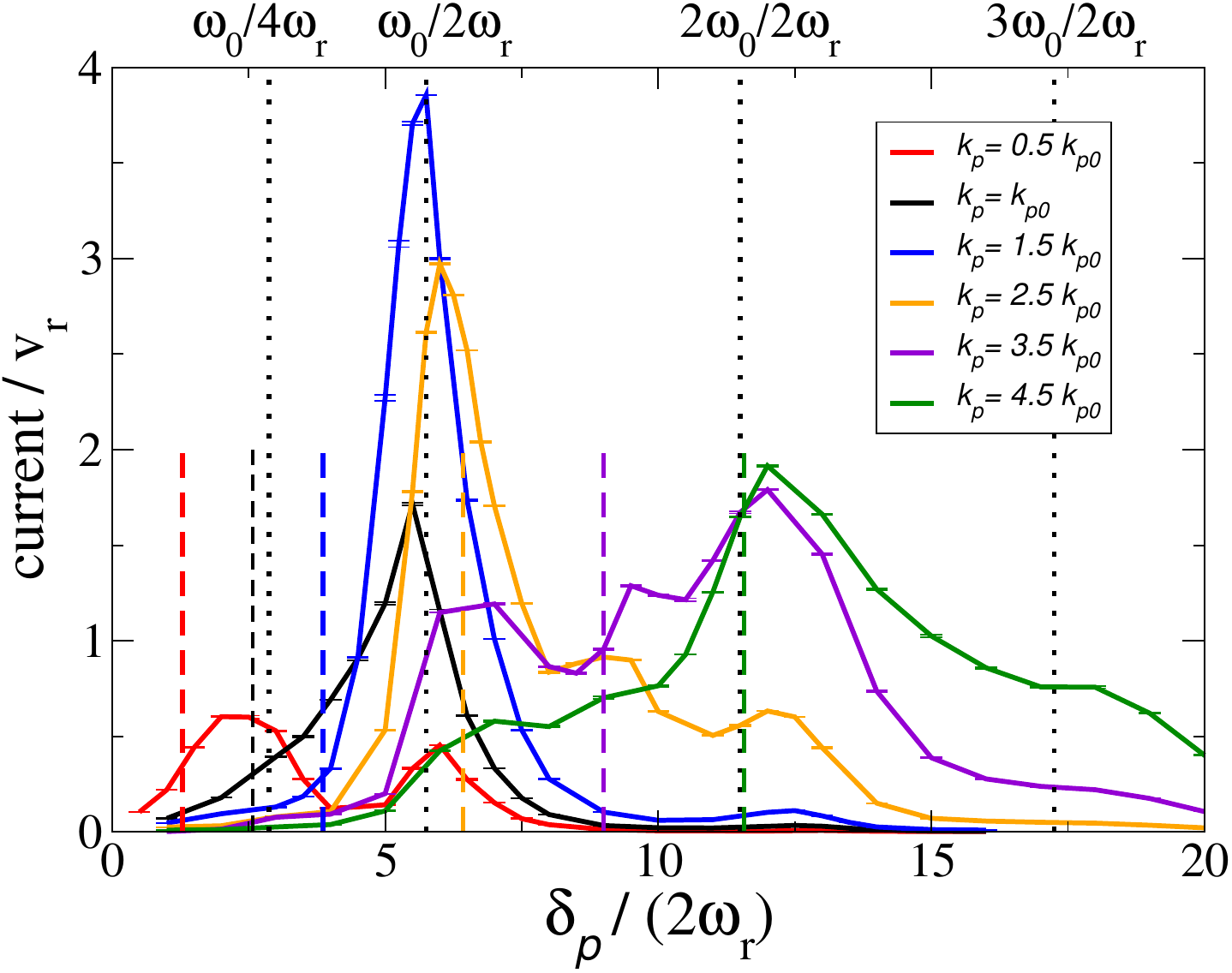}
\caption{
\textcolor{black}{Space quasiperiodic driving of the $U^{(b)}$ system in (\ref{eq:Upm:3D:b}) (i.e., case (b)): Same as in Fig.~\ref{fig:sims3}, but this time the atomic current is plotted for} space quasiperiodic driving with $k_{p0}=k_0/\sqrt{5}$. The current is plotted for several values of the driving wave number, $k_p=0.5 k_{p0}, k_{p0},1.5 k_{p0}, 2.5 k_{p0}, 3.5 k_{p0}$, and $4.5 k_{p0}$. In all cases the driving amplitude $\epsilon_p$ was varied so that  $\epsilon_p k_p/k_l=0.8$, thus keeping \textcolor{black}{fixed the value of the force amplitude $F_p$, which is defined in the text immediately following (\ref{eq:3Dlin:b})}. As in Fig.~\ref{fig:sims3}, dotted vertical lines are placed at $\omega_0/(2\omega_r)=5.75$, $\omega_0/2$, and $2\omega_0$, plus an additional one at {$3\omega_0$}.
The truncated vertical dashed lines correspond to the shifted values $(\delta_p)_{vm}=\omega_0 k_p/k_0$. 
\label{fig:sims4}
}
\end{figure}

\textcolor{black}{We plot the current in (\ref{eq:3Dlin:b}) for the $U^{(b)}$ system, as a function of $\delta_p$, in Fig.~\ref{fig:sims3} at several values of the driving amplitude $\varepsilon_p$ for the case of space periodic driving $k_p=k_0$, and in Fig.~\ref{fig:sims4} at several values of the driving wavenumber $k_p$ for the case of space quasiperiodic driving.}  

\textcolor{black}{In Fig.~\ref{fig:sims3},} a peak at $\delta_p=\omega_0=5.75$ is observed, common to all curves, except the one corresponding to the largest value of $\varepsilon_p$, which is shifted to higher frequency values.
\textcolor{black}{These peaks
are} not far from the vibrational frequency associated with linear oscillations in the optical wells{\textcolor{black}{, $\Omega_x=7.3$ (as evaluated from (\ref{eq:omx}) using $\theta_x = 25^0$, which is the value employed in our experiment). Therefore, we identify this frequency $\omega_0$ with the intrinsic vibrational frequency $\Omega_x$ of the lattice.}  
In Fig.~\ref{fig:sims3} we see that the presence of additional peaks at about $\omega_0/2$ and $2\omega_0$ \textcolor{black}{are predicted by the simulations}. This behavior is not uncommon, in rocked ratchets the current is also observed \cite{cubero2014} to peak at \textcolor{black}{multiples and sub-multiples}, in general a fractional number,
of an intrinsic frequency.

%
%


As the driving wave number $k_p$ is varied, so does the phase velocity of the propagating perturbation  {$v_p=\delta_p/k_p$}. It is expected \cite{gryn96} that a peak \textcolor{black}{in current} is produced when this phase velocity matches the velocity
\begin{equation}
 v_0=\omega_0/k_0.
 \label{eq:v0}
 \end{equation} 
 The intrinsic velocity $v_0$ is associated with an average drift in one direction due to half oscillations in a well, followed by transitions between the atomic states \cite{gryn96,proc95}.   This velocity matching ({$vm$}) mechanism  thus yields  a peak at 
\begin{equation}
 (\delta_{p})_{vm}=\omega_0 k_p / k_0.
\label{eq:deltapm}
\end{equation}

\textcolor{black}{The positions of these velocity-matched current resonances, $(\delta_{p})_{vm}$, are indicated in Fig.~\ref{fig:sims4} for six $k_p/k_0$-values, by the truncated vertical dashed lines. As explained below, the numerical results, presented in Fig.~\ref{fig:sims4}, do 
indeed predict current resonances at or near these $(\delta_{p})_{vm}$-values. However, a clear picture is obscured by the locations nearby of $\delta_{p}$-values at fractional ratios of the intrinsic vibrational frequency $\omega_0$, that were discussed above in Figs.~\ref{fig:sims3} - these frequencies are indicated by the vertical dotted lines in Figs.~\ref{fig:sims3} and Figs.~\ref{fig:sims4}.}
For example, let's consider the case of space-quasiperiodic driving with {$k_p=0.5 k_{p0}$}, where {$k_{p0} \equiv k_0/\sqrt{5}$}. In this case, the \textcolor{black}{red} current-vs-{$\delta_p$} curve \textcolor{black}{in Fig.~\ref{fig:sims4}} displays a peak near 
 ({$\delta_{p})_{vm}$} \textcolor{black}{(red truncated vertical dashed line)},
but it is also leaning towards $\omega_0/2$.  The curves for $k_p=k_{p0}$ and $k_p=1.5 k_{p0}$  show no velocity matching shift, just peaking at the value $\omega_0$. The curve  $k_p=2.5 k_{p0}$ does peak at  
 ({$\delta_{p})_{vm}$} \textcolor{black}{(yellow truncated vertical dashed line)},
but this value is very near $\omega_0$ in this case. It also shows \textcolor{black}{small peaks at about $1.5\omega_0$ and $2\omega_0$}. In the case  $k_p=3.5 k_{p0}$, the frequency 
 ({$\delta_{p})_{vm}$}
lies between the peaks at $\omega_0$ and $2\omega_0$. The corresponding curve peaks \textcolor{black}{near} $\omega_0$, 
 ({$\delta_{p})_{vm}$}
(or $1.5\omega_0$, since they are \textcolor{black}{close by} here), and  $2\omega_0$. The curve $k_p=4.5 k_{p0}$ offers a clear confirmation of the velocity matching mechanism, because it shows no clear peak at $\omega_0$, \textcolor{black}{but} a distinct one at  
 ({$\delta_{p})_{vm}$},
which \textcolor{black}{happens to practically coincide with} $2\omega_0$. 

Overall, \textcolor{black}{the simulation results in Fig.~\ref{fig:sims4} suggest that} the discussed shift due to velocity matching is clearly at play in the system, but the shift takes place through local maxima at a fractional ratio of the intrinsic frequency $\omega_0$.


We may then wonder how a particular atomic \textcolor{black}{Fourier} mode is affected by the above discussed resonances. Figure~\ref{fig:sims5} shows the \textcolor{black}{calculations} for the case $k_p=3.5 k_{p0}$, which was chosen because the current shows three peaks in Figure~\ref{fig:sims4}, at $\omega_0$, 
$1.5\omega_0\approx$ ({$\delta_{p})_{vm}$},
and $2\omega_0$.  
\begin{figure}[t]
\includegraphics[width=8cm]{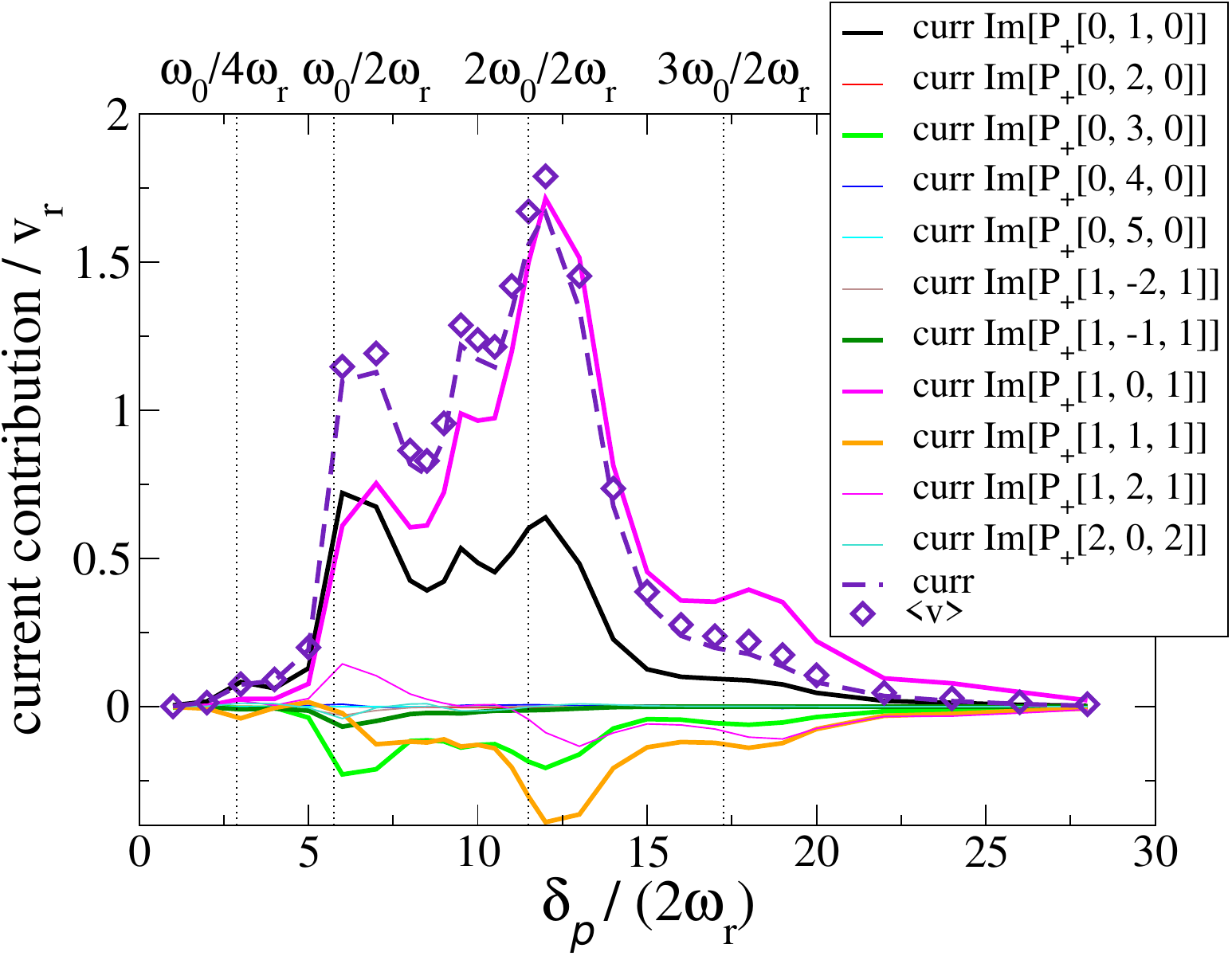}
\caption{
\textcolor{black}{Fourier mode contributions to the current} as a function of the driving frequency $\delta_p$\textcolor{black}{$=\omega_p - \omega_l$,} for the quasiperiodic case  {$k_p=3.5 k_{p0}$} shown in Fig.~\ref{fig:sims4} (for the $U^{(b)}$ system in (\ref{eq:Upm:3D:b}), case (b)). \textcolor{black}{The total current (dashed curve) is obtained by summing all the mode contributions. The diamonds represent the current calculated directly in the simulation, and their plot is identical to the $k_p = 3.5 k_{p0}$ curve in Fig.~\ref{fig:sims4}.} As in Fig.~\ref{fig:sims4}, dotted vertical lines
 \textcolor{black}{are placed at {$\omega_0/(2\omega_r)=5.75$}, {$\omega_0/2$}, {$2\omega_0$}, and $3\omega_0$.} \
\label{fig:sims5}
}
\end{figure}

The plot shows peaks at these frequency values at most of the atomic \textcolor{black}{Fourier} modes, behaving very similarly \textcolor{black}{for} all of them. Moreover, an extra peak in most of them is also visible at $\delta_p=3\omega_0$, a peak which is 
\textcolor{black}{discernable in the overall current (dashed curve). 
The good agreement between the current obtained directly in the simulations (diamonds) and the sum of all the mode contributions (dashed line) serves as a validation of the analytical calculations.}


\textcolor{black}{Looking closely at Fig.~\ref{fig:sims5} one can find subtle differences between the modes. First, it can be seen that the total current's curve follows closely the curve of the mode $[1,0,1]$, confirming this atomic mode as the dominant one. Many other modes, such as $[0,1,0]$, $[0,3,0]$, or $[1,2,1]$ curiously show a larger peak at $\omega_0$ than at $2\omega_0$. The mode $[1,1,1]$ shows a similar behavior to the dominant $[1,0,1]$, with a larger peak at $2\omega_0$, but with the current always in the opposite direction.} 

\section{Experiment}
\label{sec:expt}
Our experiments are performed 
on $^{85}$Rb atoms  
in a dilutely occupied dissipative 3D lattice 
in a tetrahedral lin$\perp$lin configuration, as in Fig.~\ref{fig:samir1}.
Directed motion is produced by the weak $\hat{y}$-polarized probe which 
makes an angle $\theta_p$ with the $z$-axis. The probe frequency $\omega_p$ is scanned around the (fixed) frequency $\omega_l$ of the lattice beams \textcolor{black}{(which collectively serve as the pump)}, and probe transmission is measured as a function of probe detuning $\delta_p = \omega_p - \omega_l$. As in Ref.~\cite{samir22}, $\delta_p/\omega_L < 10^{-9}$ \textcolor{black}{for all pump-probe spectra presented in this work}. Here, the intensity ratio of probe to lattice (sum of all four beams) is less than $3\%$.


In accordance with the notation used in Ref.~\cite{grynreview,samir22}, 
$\Delta_0^\prime \equiv \Delta {s_0}/2$, and {$\Gamma^\prime \equiv \Gamma {s_0/2}$} where $s_0 \equiv (I/I_{sat})/(1 + 4\Delta^2/\Gamma^2)$ is just the saturation parameter. For the $F_g = 3 \rightarrow F_e = 4$ transition in $^{85}$Rb, $I_{sat} = 1.67$ mW/cm$^2$ for $\sigma$-light, 
$\Gamma/2\pi$ is the natural linewidth for $^{85}$Rb (6.07 MHz), and $\omega_r/2\pi$ is the recoil frequency (3.86 kHz). In our experiment, $\theta_x   = \theta_y= 25^0$, and each lattice beam has intensity \textcolor{black}{$I = 6.22 \pm 0.22$ mW/cm$^2$}, a $1/e^2$-diameter 5.4 mm (the probe diameter is 1.4 mm), and red-detuning $\Delta = 8.75 \Gamma$.  In order to determine the intensity $I$ that actually illuminates the atoms, care is taken to account for the intensity loss through the windows of the vacuum cell that houses the lattice, and the background Rb vapor. \textcolor{black}{These values yield 
$\Gamma^\prime/2\pi = 36.8 \pm 1.3$ kHz and a well-depth $U_0 = 445 \pm 15 \,\hbar\omega_r$. Using the definition of the recoil frequency just after (\ref{eq:omx}), and setting $\hbar = m_a = k_l = 1$ as in the simulations, we find this $U_0$-value corresponds to $223 \pm 8$, and $\Gamma'$, in units of the recoil frequency, corresponds to $4.77 \pm 0.17$ - these values are comparable to those assumed in the simulations in Sec.~\ref{sec:num}. } \textcolor{black}{The pump-probe spectra measured for the probe beam propagating through the lattice cannot be directly compared to the theory curves in Figs.~\ref{fig:sims3} - \ref{fig:sims5} which plot the directed atomic currents. However, as described below, some of the observed peaks and dips in the measured spectra correspond to probe detuning values at which directed atomic transport occurs. These observed $\delta_p$-values may be profitably compared to the $\delta_p$-values in Figs.~\ref{fig:sims3} - \ref{fig:sims5} where resonant peaks in atomic current are predicted. }    


Figs.~\ref{fig:fig7}(a) and (b) show distinctly different probe transmission spectra for $\theta_p = 0$ and $17.5^0$, respectively. The peaks in the spectra correspond to photons absorbed from a lattice beam and emitted via stimulated emission into the probe, while dips correspond to photons absorbed from the probe and emitted into a lattice beam. \textcolor{black}{Note that the pump-probe spectra only show features arising from the interaction of the $\hat{y}$-polarized probe with the co-propagating $\hat{y}$-polarized lattice beams $\vec{k}_1$, $\vec{k}_2$ depicted in Fig.~\ref{fig:samir1}.
Contributions to the pump-probe spectrum from the other two lattice beams $\vec{k}_3$ and $\vec{k}_4$ are suppressed due to Doppler broadening in the $z$-direction, as pointed out earlier~\cite{aspect1998, samir22}. }

\begin{figure}[t]
\includegraphics[width=8cm]{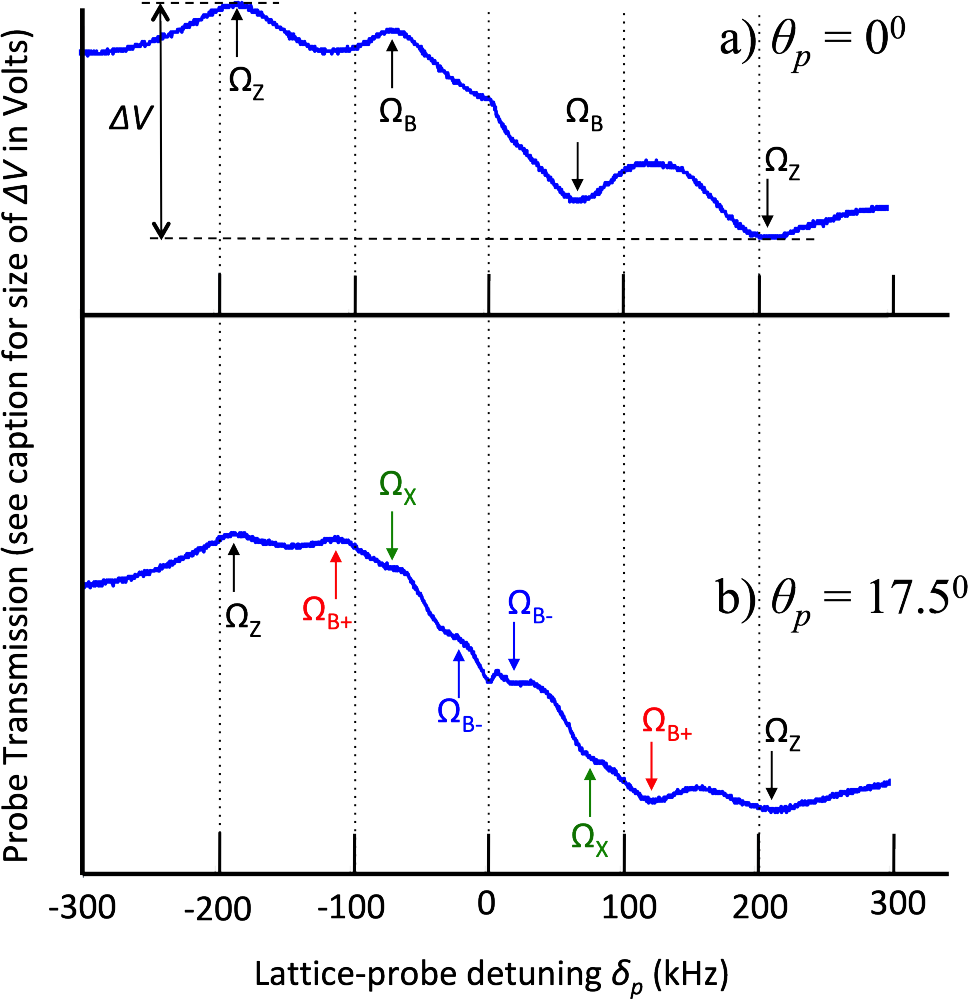}
\caption{
Probe transmission spectra measured for (a) $\theta_p = 0$ (space periodic case $k_p=k_0$) and (b) $17.5^0$ ($k_p$ not a multiple of $k_0$). 
The resonance peaks are identified as  $\Omega_{z}$, $\Omega_B$,  $\Omega_{x}$, and $\Omega_{B\pm}$ \textcolor{black}{(see text for explanation). The probe completes a frequency scan of $\pm 300$ kHz in 10 ms, generating a single-shot spectrum. The lattice is then re-loaded with atoms, a process that takes us about 5 sec, and the scan repeated. Each spectrum displayed here is an average over 16 single-shot spectra. The transmitted probe is incident on a photodetector (Thorlabs DET10A2), which is connected to an oscilloscope. The size of $\Delta V$ is typically 100 - 120 mV in our experiments. Both plots are on the same vertical scale.}
\label{fig:fig7}
}
\end{figure} 

We discuss first the periodic case shown in Fig.~\ref{fig:fig7}~(a), in which the probe is aligned with the $z$-axis ($\theta_p = 0$). This system was analyzed in detail in Ref.~\cite{samir22}, where it was confirmed that the spectral features denoted as $\Omega_z$
arise from probe-induced Raman transitions between adjacent vibrational levels corresponding to oscillations along the $z$-axis in each well, and the $\Omega_B$  features
arise from Brillouin-like directed transport along the $\pm x$-directions. 

Indeed, the vibrational frequency in the $z$-direction is determined in the system model by  \cite{samir22} 
\begin{equation}
\Omega_z  = 2(\mbox{cos}\, {\theta_x} + \mbox{cos}\,{\theta_y}) \sqrt {2|\Delta_0^\prime| \, \omega_r} ,
\end{equation}
thus yielding 
$\Omega_z  = 2\pi(180$ kHz), which is in reasonable agreement with the observed value of about 200 kHz. 
Moreover,  Eq.~(\ref{eq:omx}) yields a vibrational frequency $\Omega_x/2\pi = 60$ kHz, which is very close to the observed value for $\Omega_B/2\pi$. This $\Omega_x$-value equals 15.5 $\omega_r$, which in simulation units corresponds to 7.8, not far from the 7.3 value \textcolor{black}{calculated from (\ref{eq:omx}), as indicated in Sec.~\ref{sec:num}}. The agreement between the predicted and observed values for {$\Omega_x$} and $\Omega_B$ is remarkable considering that the theory assumed a $F_g = 1/2 \rightarrow F_e = 3/2$ atom.       

It is important to note that even though $\Omega_B$ coincides with $\Omega_x$, the spectral features at $\Omega_B$ in Fig.~\ref{fig:fig7}(a) cannot arise from intrawell oscillatory motion in the $x$-direction, owing to the fact that adjacent vibrational levels are of opposite parity, and hence the overlap integral of the probe operator (i.e., the lattice-probe interference term) between these two levels is zero. This is because the interference term goes as $\vec{E}_0 \cdot {\vec{E}^{*}}_p$ and is quadratic in $x$ for a probe that propagates purely along $\hat z$ \cite{samir22}. 

In this space periodic case $\theta_p=0$, the three probe terms of  (\ref{eq:Upm:3D:0}) reduce to a single perturbation propagating with phase velocity $v_0=\delta_p/k_0$, like in the 1D model (\ref{eq:Upm:3D:b}) with $k_p=k_0$, whose numerical results are shown in Fig.~\ref{fig:sims3}. In agreement with the experimental results of Fig.~\ref{fig:fig7}~(a), the theoretical model predicts a dominant peak at about $\omega_0$, which we identified with $\Omega_x$ in Sec.~\ref{sec:num}. The secondary peaks \textcolor{black}{predicted} at other multiples of $\omega_0$ in  Fig.~\ref{fig:sims3} for certain values of the driving amplitudes are absent in Fig.~\ref{fig:fig7}~(a).
{\color{black}This may be because the theory and simulations are performed in a 1D lattice using a $F_g = 1/2 \rightarrow F_e = 3/2$ atom with a simple ground state bipotential described by $U_\pm$ in (\ref{eq:Upm:3D:0}), whereas the experiment is performed in a 3D lattice using a $F_g = 3 \rightarrow F_e = 4$ atom, with a significantly more complex ground state manifold.} 

Let us turn our attention to the case where $k_p\ne k_0$, shown in Fig.~\ref{fig:fig7}~(b). \textcolor{black}{We see that the $\Omega_B = \Omega_x$ peak (dip) in Fig.~\ref{fig:fig7}~(a) splits into a central peak (dip) still located at $\Omega_x$, flanked by two side peaks (dips) at $\Omega_{B+}$ and $\Omega_{B-}$.} 
We have carried out several measurements with varying values of the probe angle $\theta_p$. 
\textcolor{black}{The data, shown in Fig.~\ref{fig:fig3}(a) \textcolor{black}{(only half the scan range depicted in Fig.~\ref{fig:fig7} is necessary)}, are the first time pump-probe spectra have been systematically measured in the probe-lattice geometry of Fig.~\ref{fig:samir1}(c), for several different off-axis probe angles $\theta_p$.} 
\begin{figure}[b]
\includegraphics[width=7cm]{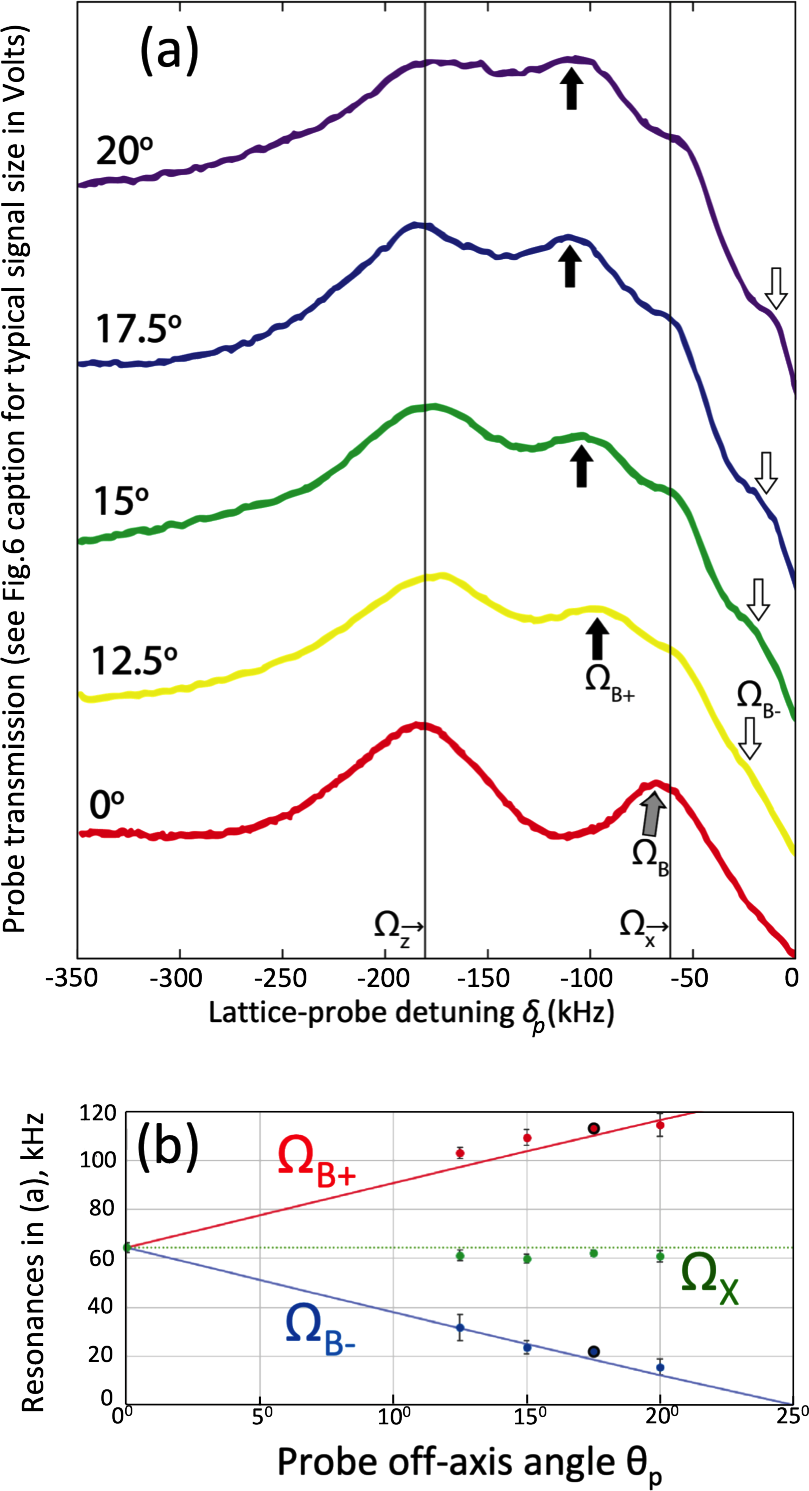}
\caption{{\hspace{0mm} 
(a) Pump-probe spectra taken at different probe angles $\theta_p$ for the lattice in Fig.~\ref{fig:samir1}(a).
The vertical lines demarcate the fixed vibrational frequencies $\Omega_z$ and $\Omega_x$. 
At $\theta_p = 0$, the gray arrow denotes the Brillouin resonance $\Omega_B$ that coincides with $\Omega_x$.  
At $\theta_p \neq 0$, the black and white arrows denote Brillouin resonances $\Omega_{B+}$ and $\Omega_{B-}$, respectively. \textcolor{black}{The curves for $\theta_p = 0^0$ and $17.5^0$ are just the $\delta_p < 0$ half of the spectra in Fig.~\ref{fig:fig7}.} (b) \textcolor{black}{From velocity-matching considerations,} $\Omega_{B+}$ and $\Omega_{B-}$ depart from $\Omega_x$ in accordance with Eqs.~(\ref{eq:ombplus}) and (\ref{eq:ombminus}), respectively: The solid lines are the theoretical predictions from the equations. \textcolor{black}{Our simulations in Sec.~\ref{sec:num} reveal that Brillouin propagation proceeds not just due to the velocity-matching considerations above, but also due to frequency-matching with the intrinsic frequency and its multiples (see text)}. The two black-circled datapoints in (b) correspond to the spectrum in Fig.~\ref{fig:fig7}(b). 
}}
\label{fig:fig3}
\end{figure} 
\textcolor{black}{In Ref.~\cite{gryn96}, for example, data is presented for just one off-axis angle.
In Ref.~\cite{renzoni03} an auxiliary pump-probe beam pair was introduced, completely separate from the lattice beams, which allowed for separation of purely lattice characteristic frequencies such as $\Omega_z$ and $\Omega_x$ from pump-probe induced effects such as $\Omega_B$ and $\Omega_{B\pm}$, however the signal-to-noise of the Brillouin spectral features is significantly reduced.}

\textcolor{black}{To extract the resonant frequencies $\Omega_{x,z}$ and $\Omega_{B\pm}$ from the spectra in Fig.~\ref{fig:fig3}(a) we initially used a fitting function:}
\begin{eqnarray}
& & A_{\Omega_Z} \mbox{exp}\left( -\frac{(\delta_p - \Omega_z)^2}{2\sigma_{\Omega_z}^2} \right) + A_{\Omega_X} \mbox{exp}\left( -\frac{(\delta_p - \Omega_x)^2}{2\sigma_{\Omega_x}^2} \right) \nonumber \\
+ & & A_{\Omega_{B+}} \mbox{exp}\left( -\frac{(\delta_p - \Omega_{B+})^2}{2\sigma_{\Omega_{B+}}^2} \right) + A_{\Omega_{B-}} \mbox{exp}\left( -\frac{(\delta_p - \Omega_{B-})^2}{2\sigma_{\Omega_{B-}}^2} \right) \nonumber \\
+ & & a_1 + a_2\delta_p \, .
\end{eqnarray}
\textcolor{black}{This function consists of a constant term, a linear term, and four Gaussian functions, one for each resonant peak in Fig.~\ref{fig:fig3}(a). When analyzing on-axis spectra ($\theta_p = 0$), the Gaussian functions corresponding to $\Omega_{B\pm}$ were removed. The above analysis works well for the on-axis spectra where the features are well-defined, but becomes less reliable when the features are less well-resolved, such as $\Omega_{B-}$ for example in the off-axis spectra (where $\theta_p \neq 0$). Therefore, we also employed a different analysis method for the off-axis spectra, which utilized the numerical first derivative of the spectra at each point. The absolute value of the derivative was taken, and the minimum value was found in the vicinity of each resonant feature.}  
 
\textcolor{black}{The $\Omega_{B+}$ resonances are stronger than $\Omega_{B-}$, so we consider those for convenience, without any loss of generality.} For the $\Omega_{B+}$ resonance, $k_p = k_0+k_l\sin\theta_p$, yielding $k_p/k_{p0}$-values corresponding to the four angles from $12.5^0$ to $20^0$ as 3.4, 3.6, 3.8, and 4.1, respectively. Note that the first two values are close to the $k_p/k_{p0}$-value of 3.5 in Fig.~\ref{fig:sims5}, which predicts peaks in the directed propagation at $\omega_0$, 
$(\delta_p)_{vm}\approx1.5\omega_0$,
and $2\omega_0$. The peak at $\omega_0 = \Omega_x \approx 2\pi\,(60 \mbox{kHz})$ is certainly observed in Fig.~\ref{fig:fig3} (a) and (b), 
as are single peaks at approximately 105 kHz and 110 kHz for $\theta_p = 12.5^0$ and $15^{0}$, respectively. The experiment does not \textcolor{black}{further resolve each of these single peaks, but these observed values lie not too far from the predicted kHz-values of 60, 90, 120. Thus the experimental findings in Fig.~\ref{fig:fig3} (a-b) and the} \textcolor{black}{detailed, rigorous} \textcolor{black}{numerical simulations in Fig.~\ref{fig:sims5} are consistent with each other}\textcolor{black}{, suggesting that the frequency-matching and velocity-matching mechanisms are equally at play.} 

 \textcolor{black}{On the other hand, if we ignore the frequency-matching mechanism, and only follow} the velocity matching argument discussed in Sec.~\ref{sec:num} \textcolor{black}{and in previous works~\cite{gryn96,proc95},} we expect a peak for the driving frequency values where the intrinsic velocity $\pm v_0$ (\ref{eq:v0}) matches the velocity of the propagating perturbation, thus leading to two possible values for the probe detuning $\delta_p$, located on either side of $\Omega_x$\textcolor{black}{~\cite{gryn96,proc95}}:
\begin{eqnarray}
\delta_1=\Omega_x \frac{\sin\theta_x +\sin\theta_p}{\sin\theta_x}, \label{eq:ombplus}  \\
\delta_2=\Omega_x \frac{\sin\theta_x -\sin\theta_p}{\sin\theta_x},   \label{eq:ombminus} 
\end{eqnarray}
where\textcolor{black}{, recall that} we have identified the vibrational frequency $\omega_0$ \textcolor{black}{with} $\Omega_x$.
The solid lines in Fig.~\ref{fig:fig3}~(b) \textcolor{black}{depict} the analytical predictions of Eqs.~(\ref{eq:ombplus}) and  (\ref{eq:ombminus}), with $\delta_1$ and $\delta_2$ \textcolor{black}{predicting} the observed peaks 
 $\Omega_{B+}$ and $\Omega_{B-}$, respectively, that are seen in  Fig.~\ref{fig:fig3}~(a) and Fig.~\ref{fig:fig7}~(b). \textcolor{black}{Thus, the predictions from arguments based on velocity-matching alone, are consistent as well with the data.}      
 
 \textcolor{black}{In other words, the locations of the resonances in the pump-probe spectra are explained by the previously known qualitative arguments based on velocity-matching alone, as well as by the detailed modeling developed here that reveals an equal role for the new mechanism of frequency-matching. We further note that the} quantitative agreement \textcolor{black}{between the data and theory - both the new detailed numerical simulations in Sec.~\ref{sec:num} and the previous qualitative arguments that focused on velocity-matching alone -} is remarkable, considering that the \textcolor{black}{theoretical arguments are} based on a 1D model with a simplified atomic transition. 

\textcolor{black}{To complete this discussion, we point out that the theory does offer} an explanation for \textcolor{black}{why} the amplitudes of the $\Omega_{B-}$ resonances are seen to be smaller than the $\Omega_{B+}$ ones~\cite{aspect1998}:  Since the spatial period $2\pi/(k_0-k_l\sin\theta_p)$ associated for the $\Omega_{B-}$ motion is larger than that of \textcolor{black}{$\Omega_{B+}$}, $2\pi/(k_0+k_l\sin\theta_p)$,  the sequence of half oscillations and well-transfers for $\Omega_{B-}$ is more likely to be interrupted by random photon recoils associated with the Sisyphus process, causing a larger damping of the directed propagation. Finally, \textcolor{black}{we comment on the peak at the intrinsic frequency $\omega_0$, seen consistently in data such as shown in} Figs.~\ref{fig:fig3}(a) and (b) at $\delta_p=\Omega_x$, regardless of the specific value of $\theta_p$\textcolor{black}{: Previously\textcolor{black}{~\cite{gryn96,proc95}}, it was pointed out that the} symmetry considerations for $\theta_p=0$ \textcolor{black}{that} forbid the appearance of a resonance in the spectrum due to localized vibrations about the well bottoms \textcolor{black}{(these considerations were} \textcolor{black}{mentioned above while discussing Fig.~\ref{fig:fig7}(a)}{\textcolor{black}{), no longer apply for} the case of $\theta_p\ne 0$. However, as shown \textcolor{black}{by the detailed numerical simulations} in Sec.~\ref{sec:num} \textcolor{black}{(see, for example, Fig.~\ref{fig:sims4})}, the observed peak \textcolor{black}{may also} \textcolor{black}{arise from} directed motion \textcolor{black}{due to frequency-matching}.

\section{Conclusions}
\label{sec:con}

We have studied the atomic waves generated in a dissipative optical lattice under a weak beam that produces a directed current which travels perpendicular to the direction of travel of the probe. 

On the theoretical side, a minimal 1D model of the experimental setup is studied in detail to elucidate the mechanisms of transport. An analytical method based on a Fourier decomposition of the current is applied to study the case when the driving potential perturbation has a spatial period which is the same as that of the underlying lattice, \textcolor{black}{the regime} of space periodic driving, and when both periods are incommensurate,  the regime of space quasiperiodic driving. 

It is numerically demonstrated that the transition between both regimes is smooth, despite the fact that the expansion for one of the probe perturbations is different in each regime. When the frequency of the probe is varied, the current, and more specifically the mode amplitudes that contribute to the current, show several peaks. One set is identified with a multiple of the intrinsic frequency $\omega_0$, and another is associated with a velocity matching mechanism, in which the velocity of the propagating modulation matches up with the average velocity of the atom in its intrawell oscillation. Both mechanisms are seen at play in both the space periodic and space quasiperiodic regime, and even at play at the same time, since the shift due to velocity matching \textcolor{black}{takes place} through local maxima at a fractional ratio of the intrinsic frequency $\omega_0$.

The pump-probe experiments confirm many of the above predictions. In the case of space-periodic driving $k_p=k_0$, with a weak beam that is aligned with the lattice symmetry axis, the probe transmission spectrum indeed reveals a dominant peak at \textcolor{black}{an intrinsic lattice frequency} $\omega_0$, 
that corresponds to a propagating Brillouin-like mode in a direction perpendicular to probe propagation. These observations are borne out by the numerical predictions in Fig.~\ref{fig:sims3}.}
In the case of 
driving with the weak beam incident at angle {$\theta_p$} relative to the lattice symmetry axis, the spectrum reveals a peak at $\omega_0$, and also additional peaks where \textcolor{black}{both the frequency-matching and} velocity matching mechanisms above are satisfied.
The angle-dependence of these additional spectral features, which correspond to two distinct propagating modulations with different spatial periods, is shown in Fig.~\ref{fig:fig3} to be in accordance with the analytical predictions. These findings are \textcolor{black}{consistent with} the numerical predictions in Figs.~\ref{fig:sims4} and~\ref{fig:sims5}, although it was not experimentally possible to tease apart the \textcolor{black}{relative} contributions to the observed resonances \textcolor{black}{arising from causes that were known previously~\cite{gryn96,proc95} 
versus those arising from the new frequency-matching condition revealed here by our detailed, rigorous numerical simulations.} 
\textcolor{black}{For example, the observed peaks near the intrinsic frequency $\omega_0 = \Omega_x = 60$ kHz for all the off-axis probe spectra in Fig.~\ref{fig:fig7}(b) and Fig.~\ref{fig:fig3} arise not only from contributions due to atoms oscillating inside localized wells as known previously, but also from atoms undergoing directed Brillouin propagation via the frequency-matching condition (see Fig.~\ref{fig:sims4}). Similarly, the observed $\Omega_{B+}$ peaks, say, in Fig.~\ref{fig:fig7}(b) and Fig.~\ref{fig:fig3}, for $\theta_p = 12.5^0$ and $15^0$ at about 105 kHz and 110 kHz, respectively, arise from contributions due to directed Brillouin propagation not only via the previously known velocity-matching mechanism in Eqn.~\ref{eq:ombplus}, but also via the frequency-matching mechanism which predicts a directed current at $2\omega_0 = 120$ kHz, close to the observed frequencies. Thus our investigation has shed new light on the origin of directed Brillouin modes propagating in a modulated dissipative optical lattice}, \textcolor{black}{and advanced our understanding of how Brownian ratcheting works in cold atom lattices.}     
We hope that the new understanding of Brillouin transport modes gained here would pave the road toward ratcheting of cold atoms confined in a weakly modulated optical lattice, along a precisely predictable, arbitrary direction~\cite{cubero2012}.


\section{Acknowledgments}
This work is supported by the Army Research Office under award/contract number W911NF2110120 and the Ministerio de Ciencia e Innovaci\'on of Spain of Spain, Grant No. PID2019-105316GB-I00 (DC).  We thank the Instrumentation Laboratory at Miami University for electronics and LabView support. We gratefully acknowledge invaluable assistance in the lab from Jordan Churi.


\appendix

\section{Full Optical potential}
\label{sec:apen0}

\textcolor{black}{
The full optical potential that is obtained when the 3D-lin$\perp$lin tetrahedral lattice is illuminated by a weak probe  laser of amplitude $E_p$ and frequency $\omega_p$, forming an angle $\theta_p$ with the $z$-axis and with its polarization parallel to the $y$-axis, is given by
\begin{eqnarray}
U_\pm(x,y,z,t)=
\frac{U_0}{2}\Big[-1 -\frac{1}{2}\cos(2 k_0 x)   -\frac{1}{2}\cos(2 k_y y) \nonumber \\
\pm  \cos(k_0 x)\cos(k_y y)\cos( k_z z)   \nonumber \\
+ \varepsilon_p\cos[-k_0 x+k_l\sin\theta_p  x +(\cos\theta_p  - \cos\theta_x ) k_l z  -\delta_p t ] \nonumber \\
+ \varepsilon_p\cos[ k_0 x+k_l\sin\theta_p  x +  (\cos\theta_p  - \cos\theta_x ) k_l z  -\delta_p t ] \nonumber \\
\pm \varepsilon_p\cos(k_y y)\cos[k_l\sin\theta_p  x -(\cos\theta_p  + \cos\theta_y ) k_l z  -\delta_p t)
\Big], \nonumber \\
\label{eq:Upm:3D}
\end{eqnarray}   
where $k_y=k_l \sin\theta_y$ and $k_z=k_l(\cos\theta_x+\cos\theta_y)/2$.
By taking $y=z=0$ in (\ref{eq:Upm:3D}), we obtain (\ref{eq:Upm:3D:0}).
}

\section{Further analytical results}
\label{sec:apen}
The setup defined by case (a) in (\ref{eq:Upm:3D:a}) requires different expressions for the quasiperiodic ({$k_p/k_0$} irrational) and periodic {$k_p=k_0$} cases. The calculation proceeds along the same lines sketched in \cite{cubero22}. The atomic state symmetry produced by the probe for setup (\ref{eq:Upm:3D:a}) is given by
\begin{equation}
P_-[l,n,m]= (-1)^{n+l} P_+[l,n,m],
\label{eq:pminus:a0}
\end{equation}
which can be also written as
\begin{equation}
P_-[l,n,m]= (-1)^{n+m} P_+[l,n,m],
\label{eq:pminus:a}
\end{equation}
the latter being a useful expression in the special (periodic) case  $k_p=k_0$.

In the quasiperiodic case ({$k_p/k_0$} irrational) we find 
\begin{eqnarray}
\langle v \rangle_\mathrm{(a)} =\langle v \rangle^{(0)}+ \frac{m_a}{m_a F_0 g_1 - 2 D_0 k_0}  \Bigg[   \nonumber \\
%
%
+\frac{\text{Im}\left[e^{i\phi_p}P_+[1,-4,1]\right] (2 F_p g_2^2 k_0) }{(2k_0-k_p) k_p}  \nonumber \\
%
+\frac{\text{Im}\left[e^{i\phi_p}P_+[1,-3,1]\right] 2 F_p g_1 g_2 k_0 }{(2k_0-k_p) k_p}  \nonumber \\
+\frac{\text{Im}\left[e^{i\phi_p}P_+[1,-2,1]\right] }{ (2k_0-k_p) k_p^2}  [8 F_p g_0 g_2 (-k_0^2  +k_0k_p) \nonumber \\
+ F_0 F_p (2 k_0^2 k_p-2k_0 k_p^2+k_p^3/2)/m_a]  \nonumber \\
-\frac{\text{Re}\left[e^{i\phi_p}P_+[1,-2,1]\right] 4 F_p g_2 k_0 (k_0-k_p)\delta_p }{(2k_0-k_p) k_p^2}  \nonumber \\
%
+\frac{\text{Im}\left[e^{i\phi_p}P_+[1,-1,1]\right] }{ (2k_0-k_p)k_p^2}  F_p \Bigg[4 g_0g_1k_0(-2k_0+k_p)   \nonumber \\ 
+2g_1g_2k_0k_p +\frac{F_0}{m_a}(-2 k_0^2 k_p +3k_0 k_p^2-k_p^3) \Bigg]   \nonumber \\ 
{\color{blue}
-\frac{\text{Re}\left[e^{i\phi_p}P_+[1,-1,1]\right] 2 F_p g_1 k_0 (k_0- 2 k_p)\delta_p }{(k_0-k_p) k_p^2} 
}
 \nonumber \\
%
-\frac{ \text{Im}\left[e^{i\phi_p}P_+[1,0,1]\right] }{(2k_0-k_p)k_p^2(2k_0+k_p)}   F_p[ g_0^2( 32 k_0^3-8 k_0k_p^2) \nonumber \\
 -4g_2^2 k_0 k_p^2 +\delta_p^2(-8k_0^3+2k_0 k_p^2) ]  \nonumber \\
-\frac{\text{Re}\left[e^{i\phi_p}P_+[1,0,1]\right] 8 F_p g_0 k_0 \delta_p }{k_p^2}  \nonumber \\
%
+\frac{\text{Im}\left[e^{i\phi_p}P_+[1,1,1]\right] }{(2k_0+k_p) k_p^2}  F_p [ -4 g_0g_1 k_0(2k_0+k_p) \nonumber \\
 -2g_1g_2k_0k_p + F_0(2k_0^2k_p+3k_0k_p^2+k_p^3)/m_a  ]  \nonumber \\
-\frac{\text{Re}\left[e^{i\phi_p}P_+[1,1,1]\right] 2 F_p g_1 k_0 (k_0+k_p)\delta_p }{(k_0+k_p) k_p^2}  \nonumber \\
%
+\frac{\text{Im}\left[e^{i\phi_p}P_+[1,2,1]\right] }{ (2k_0+k_p) k_p^2}   [8 F_p g_0 g_2 (-k_0^2-k_0k_p) \nonumber \\
+ F_0 F_p (-2 k_0^2 k_p-2k_0 k_p^2-k_p^3/2)/m_a]  \nonumber \\
-\frac{\text{Re}\left[e^{i\phi_p}P_+[1,2,1]\right] 4 F_p g_2 k_0 (k_0+k_p)\delta_p }{(2k_0+k_p) k_p^2}  \nonumber \\
%
-\frac{\text{Im}\left[e^{i\phi_p}P_+[1,3,1]\right] 2 F_p g_1 g_2 k_0  }{(2k_0+k_p) k_p}  \nonumber \\
-\frac{\text{Im}\left[e^{i\phi_p}P_+[1,4,1]\right] (2 F_p g_2^2 k_0) }{(2k_0+k_p) k_p}  \nonumber \\
%
+\frac{\text{Im}\left[e^{i2\phi_p}P_+[2,0,2]\right] F_p^2 k_0 }{m_a k_p}  
\Bigg], \label{eq:3Dlin:a:quasi}  \nonumber \\
\end{eqnarray}
where $\langle v \rangle^{(0)}$ contains no probe-related terms, and is given by
\begin{eqnarray}
\langle v \rangle^{(0)} = \frac{m_a}{m_a F_0 g_1 - 2 D_0 k_0}  \Bigg[   \nonumber \\
-\frac{\text{Im}\left[P_+[0,1,0]\right] F_0\left( 8 g_0^2 -4g_2^2/3+F_0 k_0/(2m_a)  \right)  }{k_0}  \nonumber \\
+\frac{\text{Im}\left[P_+[0,2,0]\right] F_0 \left(-4 g_0 g_1 {  -8g_1 g_2/3} +F_0 k_0/m_a\right)}{k_0}  \nonumber \\
+\frac{\text{Im}\left[P_+[0,3,0]\right] F_0 \left(-16 g_0 g_1/3 - 2 g_2^2 -3F_0 k_0/(2m_a)\right)}{k_0}  \nonumber \\
  +\frac{\text{Im}\left[P_+[0,4,0]\right] F_0 \left(-2 g_1 g_2/3 +F_0 k_0/(2m_a)\right)}{k_0}  \nonumber \\
  -\frac{\text{Im}\left[P_+[0,5,0]\right] F_0  2 g_2^2 }{3 k_0}  \Bigg]. \nonumber\\
\label{eq:v0}  
\end{eqnarray}

\textcolor{black}{The term highlighted in blue (light grey) in Eq. (\ref{eq:3Dlin:a:quasi}) is {\color{black} the contribution of the real part of the}
$[1, -1, 1]$ mode, referred to while discussing Fig.~\ref{fig:sims1} in Sec.~\ref{sec:num}.}

Equation (\ref{eq:3Dlin:a:quasi}) is not valid when the ratio $k_p/k_0$ is a rational number. In this case there are resonances which require a special derivation \cite{cubero22}. An indication of this fact is that the coefficient associated with the mode amplitude  $\mathrm{Re}[P_+[1, -1, 1]]$ goes to infinity in the limit $k_p\rightarrow k_0$, whereas the actual coefficient when computed directly in the case $k_p=k_0$ remains finite, as shown in Eq.~(\ref{eq:3Dlin:a:per}).
 
\newpage
The full expansion in the periodic case $k_p=k_0$ is given by
\begin{eqnarray}
\langle v \rangle_\mathrm{(a)} = \langle v \rangle^{(0)}
+ \frac{m_a}{m_a F_0 g_1 - 2 D_0 k_0}  \Bigg[   \nonumber \\
{\color{blue}  -\frac{\text{Re}\left[P_+[0,1,0]\right] F_0\left( F_p^2 g_1 \right)  }{m_a\delta_p}  }  
%
%
+\frac{\text{Im}\left[e^{i\phi_p}P_+[1,-4,1]\right] (2 F_p g_2^2 ) }{k_0}  \nonumber \\
{\color{blue} +\frac{\text{Re}\left[e^{i\phi_p}P_+[1,-3,1]\right] F_0 F_p g_1  }{m_a\delta_p}  }   
+\frac{\text{Im}\left[e^{i\phi_p}P_+[1,-3,1]\right] 2 F_p g_1 g_2 }{k_0}  \nonumber \\
+\frac{\text{Im}\left[e^{i\phi_p}P_+[1,-2,1]\right] F_0 F_p }{ 2m_a}   
{\color{blue} -\frac{\text{Re}\left[e^{i\phi_p}P_+[1,-2,1]\right]  F_0 F_p g_1  }{m_a \delta_p }  }\nonumber \\
+\frac{\text{Im}\left[e^{i\phi_p}P_+[1,-1,1]\right] }{ k_0}  F_p (-4 g_0 g_1  +2g_1g_2)   \nonumber \\ 
{\color{blue}  -\frac{\text{Re}\left[e^{i\phi_p}P_+[1,-1,1]\right] 2 F_p g_1 \delta_p }{k_0 }   }  \nonumber \\
-\frac{ \text{Im}\left[e^{i\phi_p}P_+[1,0,1]\right] }{3k_0}   2 F_p(12 g_0^2 -2 g_2^2-3\delta_p^2 ) \nonumber \\
-\frac{\text{Re}\left[e^{i\phi_p}P_+[1,0,1]\right] (  {\color{blue}-F_0 F_p g_1 k_0/\delta_p }   +8 F_p g_0 \delta_p ) }{k_0}  \nonumber \\   
+\frac{\text{Im}\left[e^{i\phi_p}P_+[1,1,1]\right] }{3k_0}  F_p [ -12 g_0g_1  -2g_1g_2  + 6 F_0 k_0/m_a  ]  \nonumber \\
-\frac{\text{Re}\left[e^{i\phi_p}P_+[1,1,1]\right]  {\color{blue} ( F_p F_0 g_1 k_0 /\delta_p  } + 3 F_p g_1 \delta_p ) }{k_0}  \nonumber \\
+\frac{\text{Im}\left[e^{i\phi_p}P_+[1,2,1]\right] }{ 6 k_0}  F_p  (-32 g_0 g_2 - 9 F_0 k_0/m_a )  \nonumber \\
-\frac{\text{Re}\left[e^{i\phi_p}P_+[1,2,1]\right] 8 F_p g_2   \delta_p }{ 3 k_0}  
-\frac{\text{Im}\left[e^{i\phi_p}P_+[1,3,1]\right] 2 F_p g_1 g_2  }{3 k_0}  \nonumber \\
-\frac{\text{Im}\left[e^{i\phi_p}P_+[1,4,1]\right] (2 F_p g_2^2 ) }{3 k_0}  
{\color{blue}  +\frac{\text{Im}\left[e^{i2\phi_p}P_+[2,-1,2]\right] F_p^2 g_1 }{\delta_p}  }  \nonumber \\
+\frac{\text{Im}\left[e^{i2\phi_p}P_+[2,0,2]\right] F_p^2  }{m_a}  
\Bigg] \quad (k_p=k_0). \label{eq:3Dlin:a:per}  \nonumber \\
\end{eqnarray}
Terms in (\ref{eq:3Dlin:a:per}), which are not directly obtained from (\ref{eq:3Dlin:a:quasi}) after taking the limit $k_p\rightarrow k_0$, have been highlighted in blue (light grey).  Note there are also terms in the quasiperiodic case (\ref{eq:3Dlin:a:quasi}) which are singular when $k_p\rightarrow 2k_0$. But again, this singularity is removed in the exact limit $k_p=2k_0$.


\begin{thebibliography}{40}%
\makeatletter
\providecommand \@ifxundefined [1]{%
 \@ifx{#1\undefined}
}%
\providecommand \@ifnum [1]{%
 \ifnum #1\expandafter \@firstoftwo
 \else \expandafter \@secondoftwo
 \fi
}%
\providecommand \@ifx [1]{%
 \ifx #1\expandafter \@firstoftwo
 \else \expandafter \@secondoftwo
 \fi
}%
\providecommand \natexlab [1]{#1}%
\providecommand \enquote  [1]{``#1''}%
\providecommand \bibnamefont  [1]{#1}%
\providecommand \bibfnamefont [1]{#1}%
\providecommand \citenamefont [1]{#1}%
\providecommand \href@noop [0]{\@secondoftwo}%
\providecommand \href [0]{\begingroup \@sanitize@url \@href}%
\providecommand \@href[1]{\@@startlink{#1}\@@href}%
\providecommand \@@href[1]{\endgroup#1\@@endlink}%
\providecommand \@sanitize@url [0]{\catcode `\\12\catcode `\$12\catcode
  `\&12\catcode `\#12\catcode `\^12\catcode `\_12\catcode `\%12\relax}%
\providecommand \@@startlink[1]{}%
\providecommand \@@endlink[0]{}%
\providecommand \url  [0]{\begingroup\@sanitize@url \@url }%
\providecommand \@url [1]{\endgroup\@href {#1}{\urlprefix }}%
\providecommand \urlprefix  [0]{URL }%
\providecommand \Eprint [0]{\href }%
\providecommand \doibase [0]{https://doi.org/}%
\providecommand \selectlanguage [0]{\@gobble}%
\providecommand \bibinfo  [0]{\@secondoftwo}%
\providecommand \bibfield  [0]{\@secondoftwo}%
\providecommand \translation [1]{[#1]}%
\providecommand \BibitemOpen [0]{}%
\providecommand \bibitemStop [0]{}%
\providecommand \bibitemNoStop [0]{.\EOS\space}%
\providecommand \EOS [0]{\spacefactor3000\relax}%
\providecommand \BibitemShut  [1]{\csname bibitem#1\endcsname}%
\let\auto@bib@innerbib\@empty
\bibitem [{\citenamefont {Heinrich}\ \emph {et~al.}(2021)\citenamefont
  {Heinrich}, \citenamefont {Oliver}, \citenamefont {Vandersypen},
  \citenamefont {Ardavan}, \citenamefont {Sessoli}, \citenamefont {Loss},
  \citenamefont {Jayich}, \citenamefont {Fernandez-Rossier}, \citenamefont
  {Laucht},\ and\ \citenamefont {Morello}}]{nat2021}%
  \BibitemOpen
  \bibfield  {author} {\bibinfo {author} {\bibfnamefont {A.}~\bibnamefont
  {Heinrich}}, \bibinfo {author} {\bibfnamefont {W.~D.}\ \bibnamefont
  {Oliver}}, \bibinfo {author} {\bibfnamefont {L.~M.~K.}\ \bibnamefont
  {Vandersypen}}, \bibinfo {author} {\bibfnamefont {A.}~\bibnamefont
  {Ardavan}}, \bibinfo {author} {\bibfnamefont {R.}~\bibnamefont {Sessoli}},
  \bibinfo {author} {\bibfnamefont {D.}~\bibnamefont {Loss}}, \bibinfo {author}
  {\bibfnamefont {A.~B.}\ \bibnamefont {Jayich}}, \bibinfo {author}
  {\bibfnamefont {J.}~\bibnamefont {Fernandez-Rossier}}, \bibinfo {author}
  {\bibfnamefont {A.}~\bibnamefont {Laucht}},\ and\ \bibinfo {author}
  {\bibfnamefont {A.}~\bibnamefont {Morello}},\ }\bibfield  {title} {\bibinfo
  {title} {Quantum-coherent nanoscience},\ }\href@noop {} {\bibfield  {journal}
  {\bibinfo  {journal} {Nat. Nanotechnol.}\ }\textbf {\bibinfo {volume} {16}},\
  \bibinfo {pages} {1318} (\bibinfo {year} {2021})}\BibitemShut {NoStop}%
\bibitem [{\citenamefont {Wang}\ \emph {et~al.}(2018)\citenamefont {Wang},
  \citenamefont {Subhankar}, \citenamefont {Bienias}, \citenamefont
  {\L{}\k{a}cki}, \citenamefont {Tsui}, \citenamefont {Baranov}, \citenamefont
  {Gorshkova}, \citenamefont {Zoller}, \citenamefont {Porto},\ and\
  \citenamefont {Rolston}}]{porto2018}%
  \BibitemOpen
  \bibfield  {author} {\bibinfo {author} {\bibfnamefont {Y.}~\bibnamefont
  {Wang}}, \bibinfo {author} {\bibfnamefont {S.}~\bibnamefont {Subhankar}},
  \bibinfo {author} {\bibfnamefont {P.}~\bibnamefont {Bienias}}, \bibinfo
  {author} {\bibfnamefont {M.}~\bibnamefont {\L{}\k{a}cki}}, \bibinfo {author}
  {\bibfnamefont {T.-C.}\ \bibnamefont {Tsui}}, \bibinfo {author}
  {\bibfnamefont {M.~A.}\ \bibnamefont {Baranov}}, \bibinfo {author}
  {\bibfnamefont {A.~V.}\ \bibnamefont {Gorshkova}}, \bibinfo {author}
  {\bibfnamefont {P.}~\bibnamefont {Zoller}}, \bibinfo {author} {\bibfnamefont
  {J.~V.}\ \bibnamefont {Porto}},\ and\ \bibinfo {author} {\bibfnamefont
  {S.}~\bibnamefont {Rolston}},\ }\bibfield  {title} {\bibinfo {title} {Dark
  state optical lattice with sub-wavelength spatial structure},\ }\href@noop {}
  {\bibfield  {journal} {\bibinfo  {journal} {Phys. Rev. Lett.}\ }\textbf
  {\bibinfo {volume} {120}},\ \bibinfo {pages} {083601} (\bibinfo {year}
  {2018})}\BibitemShut {NoStop}%
\bibitem [{\citenamefont {\L{}\k{a}cki}\ \emph {et~al.}(2016)\citenamefont
  {\L{}\k{a}cki}, \citenamefont {Baranov}, \citenamefont {Pichler}, ,\ and\
  \citenamefont {Zoller}}]{zoller2016}%
  \BibitemOpen
  \bibfield  {author} {\bibinfo {author} {\bibfnamefont {M.}~\bibnamefont
  {\L{}\k{a}cki}}, \bibinfo {author} {\bibfnamefont {M.~A.}\ \bibnamefont
  {Baranov}}, \bibinfo {author} {\bibfnamefont {H.}~\bibnamefont {Pichler}}, ,\
  and\ \bibinfo {author} {\bibfnamefont {P.}~\bibnamefont {Zoller}},\
  }\bibfield  {title} {\bibinfo {title} {Nanoscale dark state optical
  potentials for cold atoms},\ }\href@noop {} {\bibfield  {journal} {\bibinfo
  {journal} {Phys. Rev. Lett.}\ }\textbf {\bibinfo {volume} {117}},\ \bibinfo
  {pages} {233001} (\bibinfo {year} {2016})}\BibitemShut {NoStop}%
\bibitem [{\citenamefont {Miles}\ \emph {et~al.}(2013)\citenamefont {Miles},
  \citenamefont {Simmons},\ and\ \citenamefont {Yavuz}}]{yavuz2013}%
  \BibitemOpen
  \bibfield  {author} {\bibinfo {author} {\bibfnamefont {J.~A.}\ \bibnamefont
  {Miles}}, \bibinfo {author} {\bibfnamefont {Z.~J.}\ \bibnamefont {Simmons}},\
  and\ \bibinfo {author} {\bibfnamefont {D.~D.}\ \bibnamefont {Yavuz}},\
  }\bibfield  {title} {\bibinfo {title} {Subwavelength localization of atomic
  excitation using electromagnetically induced transparency},\ }\href@noop {}
  {\bibfield  {journal} {\bibinfo  {journal} {Phys. Rev. X}\ }\textbf {\bibinfo
  {volume} {3}},\ \bibinfo {pages} {031014} (\bibinfo {year}
  {2013})}\BibitemShut {NoStop}%
\bibitem [{\citenamefont {Zemanek}\ \emph {et~al.}(2020)\citenamefont
  {Zemanek}, \citenamefont {Volpe}, \citenamefont {Jonas},\ and\ \citenamefont
  {Brzobohaty}}]{optica2020}%
  \BibitemOpen
  \bibfield  {author} {\bibinfo {author} {\bibfnamefont {P.}~\bibnamefont
  {Zemanek}}, \bibinfo {author} {\bibfnamefont {G.}~\bibnamefont {Volpe}},
  \bibinfo {author} {\bibfnamefont {A.}~\bibnamefont {Jonas}},\ and\ \bibinfo
  {author} {\bibfnamefont {O.}~\bibnamefont {Brzobohaty}},\ }\bibfield  {title}
  {\bibinfo {title} {Perspective on light-induced transport of particles: from
  optical forces to phoretic motion},\ }\href@noop {} {\bibfield  {journal}
  {\bibinfo  {journal} {Adv. Opt. Photonics}\ }\textbf {\bibinfo {volume}
  {11}},\ \bibinfo {pages} {577} (\bibinfo {year} {2020})}\BibitemShut
  {NoStop}%
\bibitem [{\citenamefont {Cubero}\ and\ \citenamefont
  {Renzoni}(2016)}]{cubren16}%
  \BibitemOpen
  \bibfield  {author} {\bibinfo {author} {\bibfnamefont {D.}~\bibnamefont
  {Cubero}}\ and\ \bibinfo {author} {\bibfnamefont {F.}~\bibnamefont
  {Renzoni}},\ }\href@noop {} {{\it {\bibinfo {title} {Brownian Ratchets: From
  Statistical Physics to Bio and Nano-motors}}}}\ (\bibinfo  {publisher}
  {Cambridge University Press},\ \bibinfo {address} {Cambridge},\ \bibinfo
  {year} {2016})\BibitemShut {NoStop}%
  \bibitem [{\citenamefont {H$\ddot{\mbox{a}}$nggi}\ and\ \citenamefont
  {Marchesoni}(2009)}]{hanggi2009}%
   \BibitemOpen
  \bibfield  {author} {\bibinfo {author} {\bibfnamefont {P.}~\bibnamefont
  {H$\ddot{\mbox{a}}$nggi}} \ and\ \bibinfo {author} {\bibfnamefont {F.}~\bibnamefont
  {Marchesoni}}, \bibfield  {title} {\bibinfo {title} {Artificial Brownian motors: Controlling transport on the nanoscale}},\
  }\href@noop {} {\bibfield  {journal} {\bibinfo  {journal} {Rev. Mod. Phys.}\
  }\textbf {\bibinfo {volume} {81}},\ \bibinfo {pages} {387} (\bibinfo
  {year} {2009})}\BibitemShut {NoStop}%
\bibitem [{\citenamefont {Reimann}}(2002)]{reimann2002}%
    \BibitemOpen
  \bibfield  {author} {\bibinfo {author} {\bibfnamefont {P.}~\bibnamefont
  {Reimann}}, \bibfield  {title} {\bibinfo {title} {Brownian motors: Noisy transport far from equilibrium}},\
  }\href@noop {} {\bibfield  {journal} {\bibinfo  {journal} {Phys. Rep.}\
  }\textbf {\bibinfo {volume} {361}},\ \bibinfo {pages} {57} (\bibinfo
  {year} {2002})}\BibitemShut {NoStop}%
\bibitem [{\citenamefont {Cubero}\ and\ \citenamefont
  {Renzoni}(2012)}]{cubero2012}%
  \BibitemOpen
  \bibfield  {author} {\bibinfo {author} {\bibfnamefont {D.}~\bibnamefont
  {Cubero}}\ and\ \bibinfo {author} {\bibfnamefont {F.}~\bibnamefont
  {Renzoni}},\ }\bibfield  {title} {\bibinfo {title} {Control of transport in
  two-dimensional systems via dynamical decoupling of degrees of freedom with
  quasiperiodic driving fields},\ }\href@noop {} {\bibfield  {journal}
  {\bibinfo  {journal} {Phys. Rev. E}\ }\textbf {\bibinfo {volume} {86}},\
  \bibinfo {pages} {056201} (\bibinfo {year} {2012})}\BibitemShut {NoStop}%
\bibitem [{\citenamefont {Courtois}\ \emph {et~al.}(1996)\citenamefont
  {Courtois}, \citenamefont {Guibal}, \citenamefont {Meacher}, \citenamefont
  {Verkerk},\ and\ \citenamefont {Grynberg}}]{gryn96}%
  \BibitemOpen
  \bibfield  {author} {\bibinfo {author} {\bibfnamefont {J.~Y.}\ \bibnamefont
  {Courtois}}, \bibinfo {author} {\bibfnamefont {S.}~\bibnamefont {Guibal}},
  \bibinfo {author} {\bibfnamefont {D.~R.}\ \bibnamefont {Meacher}}, \bibinfo
  {author} {\bibfnamefont {P.}~\bibnamefont {Verkerk}},\ and\ \bibinfo {author}
  {\bibfnamefont {G.}~\bibnamefont {Grynberg}},\ }\bibfield  {title} {\bibinfo
  {title} {Propagating elementary excitation in a dilute optical lattice},\
  }\href@noop {} {\bibfield  {journal} {\bibinfo  {journal} {Phys. Rev. Lett.}\
  }\textbf {\bibinfo {volume} {77}},\ \bibinfo {pages} {40} (\bibinfo {year}
  {1996})}\BibitemShut {NoStop}%
 \bibitem [{\citenamefont {Courtois}\ (1995)\citenamefont
  {Courtois}}]{proc95}%
  \bibfield  {author} {\bibinfo {author} {\bibfnamefont {J.~Y.}\ \bibnamefont
  {Courtois}}, \ }\bibfield  {title} {\bibinfo
  {title} {Coherent and Collective Interactions of Particles and Radiation Beams},\
  }\href@noop {} {\bibfield  {journal} {\bibinfo  {journal} {Proceedings of the 
  International School of Physics ``Enrico Fermi", Course CXXXI, edited by 
  A. Aspect, W. Barletta, and R. Bonifacio}}
  (\bibinfo {year}
  {1995})}\BibitemShut {NoStop}%
  \bibitem [{\citenamefont {Jurczak}\ \emph {et~al.}(1998)\citenamefont
  {Jurczak}, \citenamefont {Courtois}, \citenamefont {Desruelle}, \citenamefont
  {Westbrook}, \ and\ \citenamefont {Aspect}}]{aspect1998}%
  \BibitemOpen
  \bibfield  {author} {\bibinfo {author} {\bibfnamefont {C.}~\bibnamefont
  {Jurczak}}, \bibinfo {author} {\bibfnamefont {J.-Y.}\ \bibnamefont
  {Courtois}}, \bibinfo {author} {\bibfnamefont {B.}~\bibnamefont {Desruelle}},
  \bibinfo {author} {\bibfnamefont {C.}~\bibnamefont {Westbrook}}, \ and\
  \bibinfo {author} {\bibfnamefont {A.}~\bibnamefont {Aspect}},\ }\bibfield
  {title} {\bibinfo {title} {Spontaneous light scattering from propagating
  density fluctuations in an optical lattice},\ }\href@noop {} {\bibfield
  {journal} {\bibinfo  {journal} {Eur. Phys. J. D}\ }\textbf {\bibinfo {volume}
  {1}},\ \bibinfo {pages} {53} (\bibinfo {year} {1998})}\BibitemShut {NoStop}%
\bibitem [{\citenamefont {Mennerat-Robilliard}\ \emph
  {et~al.}(1999)\citenamefont {Mennerat-Robilliard}, \citenamefont {Lucas},
  \citenamefont {Guibal}, \citenamefont {Tabosa}, \citenamefont {Jurczak},
  \citenamefont {Courtois},\ and\ \citenamefont {Grynberg}}]{cecile99}%
  \BibitemOpen
  \bibfield  {author} {\bibinfo {author} {\bibfnamefont {C.}~\bibnamefont
  {Mennerat-Robilliard}}, \bibinfo {author} {\bibfnamefont {D.}~\bibnamefont
  {Lucas}}, \bibinfo {author} {\bibfnamefont {S.}~\bibnamefont {Guibal}},
  \bibinfo {author} {\bibfnamefont {J.}~\bibnamefont {Tabosa}}, \bibinfo
  {author} {\bibfnamefont {C.}~\bibnamefont {Jurczak}}, \bibinfo {author}
  {\bibfnamefont {J.-Y.}\ \bibnamefont {Courtois}},\ and\ \bibinfo {author}
  {\bibfnamefont {G.}~\bibnamefont {Grynberg}},\ }\bibfield  {title} {\bibinfo
  {title} {Ratchet for cold rubidium atoms: The asymmetric optical lattice},\
  }\href@noop {} {\bibfield  {journal} {\bibinfo  {journal} {Phys. Rev. Lett.}\
  }\textbf {\bibinfo {volume} {82}},\ \bibinfo {pages} {851} (\bibinfo {year}
  {1999})}\BibitemShut {NoStop}%
\bibitem [{\citenamefont {Grynberg}\ and\ \citenamefont
  {Robilliard}(2001)}]{grynreview}%
  \BibitemOpen
  \bibfield  {author} {\bibinfo {author} {\bibfnamefont {G.}~\bibnamefont
  {Grynberg}}\ and\ \bibinfo {author} {\bibfnamefont {C.}~\bibnamefont
  {Robilliard}},\ }\bibfield  {title} {\bibinfo {title} {Cold atoms in
  dissipative optical lattices},\ }\href@noop {} {\bibfield  {journal}
  {\bibinfo  {journal} {Phys. Rep.}\ }\textbf {\bibinfo {volume} {355}},\
  \bibinfo {pages} {335} (\bibinfo {year} {2001})}\BibitemShut {NoStop}%
\bibitem [{\citenamefont {Schiavoni}\ \emph
  {et~al.}(2002{\natexlab{a}})\citenamefont {Schiavoni}, \citenamefont
  {Sanchez-Palencia}, \citenamefont {Carminati}, \citenamefont {Renzoni},\ and\
  \citenamefont {Grynberg}}]{schiren02}%
  \BibitemOpen
  \bibfield  {author} {\bibinfo {author} {\bibfnamefont {M.}~\bibnamefont
  {Schiavoni}}, \bibinfo {author} {\bibfnamefont {L.}~\bibnamefont
  {Sanchez-Palencia}}, \bibinfo {author} {\bibfnamefont {F.}~\bibnamefont
  {Carminati}}, \bibinfo {author} {\bibfnamefont {F.}~\bibnamefont {Renzoni}},\
  and\ \bibinfo {author} {\bibfnamefont {G.}~\bibnamefont {Grynberg}},\
  }\bibfield  {title} {\bibinfo {title} {Dark propagation modes in optical
  lattices},\ }\href@noop {} {\bibfield  {journal} {\bibinfo  {journal} {Phys.
  Rev. A}\ }\textbf {\bibinfo {volume} {66}},\ \bibinfo {pages} {053821}
  (\bibinfo {year} {2002}{\natexlab{a}})}\BibitemShut {NoStop}%
\bibitem [{\citenamefont {Sanchez-Palencia}\ \emph {et~al.}(2002)\citenamefont
  {Sanchez-Palencia}, \citenamefont {Carminati}, \citenamefont {Schiavoni},
  \citenamefont {Renzoni},\ and\ \citenamefont {Grynberg}}]{renzoni02a}%
  \BibitemOpen
  \bibfield  {author} {\bibinfo {author} {\bibfnamefont {L.}~\bibnamefont
  {Sanchez-Palencia}}, \bibinfo {author} {\bibfnamefont {F.-R.}\ \bibnamefont
  {Carminati}}, \bibinfo {author} {\bibfnamefont {M.}~\bibnamefont
  {Schiavoni}}, \bibinfo {author} {\bibfnamefont {F.}~\bibnamefont {Renzoni}},\
  and\ \bibinfo {author} {\bibfnamefont {G.}~\bibnamefont {Grynberg}},\
  }\bibfield  {title} {\bibinfo {title} {Brillouin propagation modes in optical
  lattices: Interpretation in terms of nonconventional stochastic resonance},\
  }\href@noop {} {\bibfield  {journal} {\bibinfo  {journal} {Phys. Rev. Lett.}\
  }\textbf {\bibinfo {volume} {88}},\ \bibinfo {pages} {133903} (\bibinfo
  {year} {2002})}\BibitemShut {NoStop}%
\bibitem [{\citenamefont {Carminati}\ \emph {et~al.}(2003)\citenamefont
  {Carminati}, \citenamefont {Schiavoni}, \citenamefont {Todorov},
  \citenamefont {Renzoni},\ and\ \citenamefont {Grynberg}}]{renzoni03}%
  \BibitemOpen
  \bibfield  {author} {\bibinfo {author} {\bibfnamefont {F.-R.}\ \bibnamefont
  {Carminati}}, \bibinfo {author} {\bibfnamefont {M.}~\bibnamefont
  {Schiavoni}}, \bibinfo {author} {\bibfnamefont {Y.}~\bibnamefont {Todorov}},
  \bibinfo {author} {\bibfnamefont {F.}~\bibnamefont {Renzoni}},\ and\ \bibinfo
  {author} {\bibfnamefont {G.}~\bibnamefont {Grynberg}},\ }\bibfield  {title}
  {\bibinfo {title} {Pump-probe spectroscopy of atoms cooled in a 3d
  lin-perp-lin optical lattice},\ }\href@noop {} {\bibfield  {journal}
  {\bibinfo  {journal} {Eur. Phys. J. D}\ }\textbf {\bibinfo {volume} {22}},\
  \bibinfo {pages} {311} (\bibinfo {year} {2003})}\BibitemShut {NoStop}%
\bibitem [{\citenamefont {Sanchez-Palencia}\ and\ \citenamefont
  {Grynberg}(2003)}]{grynpra2003}%
  \BibitemOpen
  \bibfield  {author} {\bibinfo {author} {\bibfnamefont {L.}~\bibnamefont
  {Sanchez-Palencia}}\ and\ \bibinfo {author} {\bibfnamefont {G.}~\bibnamefont
  {Grynberg}},\ }\bibfield  {title} {\bibinfo {title} {Synchronization of
  hamiltonian motion and dissipative effects in optical lattices: Evidence for
  a stochastic resonance},\ }\href@noop {} {\bibfield  {journal} {\bibinfo
  {journal} {Phys. Rev. A}\ }\textbf {\bibinfo {volume} {68}},\ \bibinfo
  {pages} {023404} (\bibinfo {year} {2003})}\BibitemShut {NoStop}%
\bibitem [{\citenamefont {Schiavoni}\ \emph {et~al.}(2003)\citenamefont
  {Schiavoni}, \citenamefont {Sanchez-Palencia}, \citenamefont {Renzoni},\ and\
  \citenamefont {Grynberg}}]{schiavoni2003}%
  \BibitemOpen
  \bibfield  {author} {\bibinfo {author} {\bibfnamefont {M.}~\bibnamefont
  {Schiavoni}}, \bibinfo {author} {\bibfnamefont {L.}~\bibnamefont
  {Sanchez-Palencia}}, \bibinfo {author} {\bibfnamefont {F.}~\bibnamefont
  {Renzoni}},\ and\ \bibinfo {author} {\bibfnamefont {G.}~\bibnamefont
  {Grynberg}},\ }\bibfield  {title} {\bibinfo {title} {Phase control of
  directed diffusion in a symmetric optical lattice},\ }\href@noop {}
  {\bibfield  {journal} {\bibinfo  {journal} {Phys. Rev. Lett.}\ }\textbf
  {\bibinfo {volume} {90}},\ \bibinfo {pages} {094101} (\bibinfo {year}
  {2003})}\BibitemShut {NoStop}%
\bibitem [{\citenamefont {Jones}\ \emph {et~al.}(2004)\citenamefont {Jones},
  \citenamefont {Goonasekara},\ and\ \citenamefont {Renzoni}}]{renzoniprl2004}%
  \BibitemOpen
  \bibfield  {author} {\bibinfo {author} {\bibfnamefont {P.~H.}\ \bibnamefont
  {Jones}}, \bibinfo {author} {\bibfnamefont {M.}~\bibnamefont {Goonasekara}},\
  and\ \bibinfo {author} {\bibfnamefont {F.}~\bibnamefont {Renzoni}},\
  }\bibfield  {title} {\bibinfo {title} {Rectifying fluctuations in an optical
  lattice},\ }\href@noop {} {\bibfield  {journal} {\bibinfo  {journal} {Phys.
  Rev. Lett.}\ }\textbf {\bibinfo {volume} {93}},\ \bibinfo {pages} {073904}
  (\bibinfo {year} {2004})}\BibitemShut {NoStop}%
\bibitem [{\citenamefont {Gommers}\ \emph {et~al.}(2005)\citenamefont
  {Gommers}, \citenamefont {Bergamini},\ and\ \citenamefont
  {Renzoni}}]{gommers2005}%
  \BibitemOpen
  \bibfield  {author} {\bibinfo {author} {\bibfnamefont {R.}~\bibnamefont
  {Gommers}}, \bibinfo {author} {\bibfnamefont {S.}~\bibnamefont {Bergamini}},\
  and\ \bibinfo {author} {\bibfnamefont {F.}~\bibnamefont {Renzoni}},\
  }\bibfield  {title} {\bibinfo {title} {Dissipation-induced symmetry breaking
  in a driven optical lattice},\ }\href@noop {} {\bibfield  {journal} {\bibinfo
   {journal} {Phys. Rev. Lett.}\ }\textbf {\bibinfo {volume} {95}},\ \bibinfo
  {pages} {073003} (\bibinfo {year} {2005})}\BibitemShut {NoStop}%
\bibitem [{\citenamefont {Gommers}\ \emph {et~al.}(2006)\citenamefont
  {Gommers}, \citenamefont {Denisov},\ and\ \citenamefont
  {Renzoni}}]{gommers2006}%
  \BibitemOpen
  \bibfield  {author} {\bibinfo {author} {\bibfnamefont {R.}~\bibnamefont
  {Gommers}}, \bibinfo {author} {\bibfnamefont {S.}~\bibnamefont {Denisov}},\
  and\ \bibinfo {author} {\bibfnamefont {F.}~\bibnamefont {Renzoni}},\
  }\bibfield  {title} {\bibinfo {title} {Quasiperiodically driven ratchets for
  cold atoms},\ }\href@noop {} {\bibfield  {journal} {\bibinfo  {journal}
  {Phys. Rev. Lett.}\ }\textbf {\bibinfo {volume} {96}},\ \bibinfo {pages}
  {240604} (\bibinfo {year} {2006})}\BibitemShut {NoStop}%
\bibitem [{\citenamefont {Gommers}\ \emph {et~al.}(2007)\citenamefont
  {Gommers}, \citenamefont {Brown},\ and\ \citenamefont
  {Renzoni}}]{renzoni2007pra}%
  \BibitemOpen
  \bibfield  {author} {\bibinfo {author} {\bibfnamefont {R.}~\bibnamefont
  {Gommers}}, \bibinfo {author} {\bibfnamefont {M.}~\bibnamefont {Brown}},\
  and\ \bibinfo {author} {\bibfnamefont {F.}~\bibnamefont {Renzoni}},\
  }\bibfield  {title} {\bibinfo {title} {Symmetry and transport in a cold atom
  ratchet with multifrequency driving,},\ }\href@noop {} {\bibfield  {journal}
  {\bibinfo  {journal} {Phys. Rev. A}\ }\textbf {\bibinfo {volume} {75}},\
  \bibinfo {pages} {053406} (\bibinfo {year} {2007})}\BibitemShut {NoStop}%
\bibitem [{\citenamefont {Gommers}\ \emph {et~al.}(2008)\citenamefont
  {Gommers}, \citenamefont {andM. Brown},\ and\ \citenamefont
  {Renzoni}}]{renzoni2008prl}%
  \BibitemOpen
  \bibfield  {author} {\bibinfo {author} {\bibfnamefont {R.}~\bibnamefont
  {Gommers}}, \bibinfo {author} {\bibfnamefont {V.~L.}\ \bibnamefont {andM.
  Brown}},\ and\ \bibinfo {author} {\bibfnamefont {F.}~\bibnamefont
  {Renzoni}},\ }\bibfield  {title} {\bibinfo {title} {Gating ratchet for cold
  atoms},\ }\href@noop {} {\bibfield  {journal} {\bibinfo  {journal} {Phys.
  Rev. Lett.}\ }\textbf {\bibinfo {volume} {100}},\ \bibinfo {pages} {040603}
  (\bibinfo {year} {2008})}\BibitemShut {NoStop}%
\bibitem [{\citenamefont {Lebedev}\ and\ \citenamefont
  {Renzoni}(2009)}]{lebedev2009}%
  \BibitemOpen
  \bibfield  {author} {\bibinfo {author} {\bibfnamefont {V.}~\bibnamefont
  {Lebedev}}\ and\ \bibinfo {author} {\bibfnamefont {F.}~\bibnamefont
  {Renzoni}},\ }\bibfield  {title} {\bibinfo {title} {Two-dimensional rocking
  ratchet for cold atoms},\ }\href@noop {} {\bibfield  {journal} {\bibinfo
  {journal} {Phys. Rev. A}\ }\textbf {\bibinfo {volume} {80}},\ \bibinfo
  {pages} {023422} (\bibinfo {year} {2009})}\BibitemShut {NoStop}%
\bibitem [{\citenamefont {Cubero}\ \emph {et~al.}(2010)\citenamefont {Cubero},
  \citenamefont {Lebedev},\ and\ \citenamefont {Renzoni}}]{cubero2010}%
  \BibitemOpen
  \bibfield  {author} {\bibinfo {author} {\bibfnamefont {D.}~\bibnamefont
  {Cubero}}, \bibinfo {author} {\bibfnamefont {V.}~\bibnamefont {Lebedev}},\
  and\ \bibinfo {author} {\bibfnamefont {F.}~\bibnamefont {Renzoni}},\
  }\bibfield  {title} {\bibinfo {title} {Current reversals in a rocking
  ratchet: dynamical vs symmetry-breaking mechanisms},\ }\href@noop {}
  {\bibfield  {journal} {\bibinfo  {journal} {Phys. Rev. E}\ }\textbf {\bibinfo
  {volume} {82}},\ \bibinfo {pages} {041116} (\bibinfo {year}
  {2010})}\BibitemShut {NoStop}%
\bibitem [{\citenamefont {Wickenbrock}\ \emph {et~al.}(2011)\citenamefont
  {Wickenbrock}, \citenamefont {Cubero}, \citenamefont {Wahab}, \citenamefont
  {Phoonthong},\ and\ \citenamefont {Renzoni}}]{cubero2011pre}%
  \BibitemOpen
  \bibfield  {author} {\bibinfo {author} {\bibfnamefont {A.}~\bibnamefont
  {Wickenbrock}}, \bibinfo {author} {\bibfnamefont {D.}~\bibnamefont {Cubero}},
  \bibinfo {author} {\bibfnamefont {N.~A.~A.}\ \bibnamefont {Wahab}}, \bibinfo
  {author} {\bibfnamefont {P.}~\bibnamefont {Phoonthong}},\ and\ \bibinfo
  {author} {\bibfnamefont {F.}~\bibnamefont {Renzoni}},\ }\bibfield  {title}
  {\bibinfo {title} {Current reversals in a rocking ratchet: The frequency
  domain},\ }\href@noop {} {\bibfield  {journal} {\bibinfo  {journal} {Phys.
  Rev. E}\ }\textbf {\bibinfo {volume} {84}},\ \bibinfo {pages} {021127}
  (\bibinfo {year} {2011})}\BibitemShut {NoStop}%
\bibitem [{\citenamefont {Hagman}\ \emph {et~al.}(2011)\citenamefont {Hagman},
  \citenamefont {Zelan}, \citenamefont {Dion},\ and\ \citenamefont
  {Kastberg}}]{umeapra12011}%
  \BibitemOpen
  \bibfield  {author} {\bibinfo {author} {\bibfnamefont {H.}~\bibnamefont
  {Hagman}}, \bibinfo {author} {\bibfnamefont {M.}~\bibnamefont {Zelan}},
  \bibinfo {author} {\bibfnamefont {C.~M.}\ \bibnamefont {Dion}},\ and\
  \bibinfo {author} {\bibfnamefont {A.}~\bibnamefont {Kastberg}},\ }\bibfield
  {title} {\bibinfo {title} {Directed transport with real-time steering and
  drifts along predesigned paths using a brownian motor,},\ }\href@noop {}
  {\bibfield  {journal} {\bibinfo  {journal} {Phys. Rev. A}\ }\textbf {\bibinfo
  {volume} {83}},\ \bibinfo {pages} {020101(R)} (\bibinfo {year}
  {2011})}\BibitemShut {NoStop}%
\bibitem [{\citenamefont {Zelan}\ \emph {et~al.}(2011)\citenamefont {Zelan},
  \citenamefont {Hagman}, \citenamefont {Labaigt}, \citenamefont {Jonsell}, ,\
  and\ \citenamefont {Dion}}]{umeapra22011}%
  \BibitemOpen
  \bibfield  {author} {\bibinfo {author} {\bibfnamefont {M.}~\bibnamefont
  {Zelan}}, \bibinfo {author} {\bibfnamefont {H.}~\bibnamefont {Hagman}},
  \bibinfo {author} {\bibfnamefont {G.}~\bibnamefont {Labaigt}}, \bibinfo
  {author} {\bibfnamefont {S.}~\bibnamefont {Jonsell}}, ,\ and\ \bibinfo
  {author} {\bibfnamefont {C.~M.}\ \bibnamefont {Dion}},\ }\bibfield  {title}
  {\bibinfo {title} {Experimental measurement of efficiency and transport
  coherence of a cold-atom brownian motor in optical lattices},\ }\href@noop {}
  {\bibfield  {journal} {\bibinfo  {journal} {Phys. Rev. A}\ }\textbf {\bibinfo
  {volume} {83}},\ \bibinfo {pages} {020102(R)} (\bibinfo {year}
  {2011})}\BibitemShut {NoStop}%
\bibitem [{\citenamefont {Wickenbrock}\ \emph {et~al.}(2012)\citenamefont
  {Wickenbrock}, \citenamefont {Holz}, \citenamefont {Wahab}, \citenamefont
  {Phoonthong}, \citenamefont {Cubero},\ and\ \citenamefont
  {Renzoni}}]{wichol12}%
  \BibitemOpen
  \bibfield  {author} {\bibinfo {author} {\bibfnamefont {A.}~\bibnamefont
  {Wickenbrock}}, \bibinfo {author} {\bibfnamefont {P.~C.}\ \bibnamefont
  {Holz}}, \bibinfo {author} {\bibfnamefont {N.~A.~A.}\ \bibnamefont {Wahab}},
  \bibinfo {author} {\bibfnamefont {P.}~\bibnamefont {Phoonthong}}, \bibinfo
  {author} {\bibfnamefont {D.}~\bibnamefont {Cubero}},\ and\ \bibinfo {author}
  {\bibfnamefont {F.}~\bibnamefont {Renzoni}},\ }\bibfield  {title} {\bibinfo
  {title} {Vibrational mechanics in an optical lattice: Controlling transport
  via potential renormalization},\ }\href@noop {} {\bibfield  {journal}
  {\bibinfo  {journal} {Phys. Rev. Lett.}\ }\textbf {\bibinfo {volume} {108}},\
  \bibinfo {pages} {020603} (\bibinfo {year} {2012})}\BibitemShut {NoStop}%
\bibitem [{\citenamefont {Staron}\ \emph {et~al.}(2022)\citenamefont {Staron},
  \citenamefont {Jiang}, \citenamefont {Scoggins}, \citenamefont {Wingert},
  \citenamefont {Cubero},\ and\ \citenamefont {Bali}}]{samir22}%
  \BibitemOpen
  \bibfield  {author} {\bibinfo {author} {\bibfnamefont {A.}~\bibnamefont
  {Staron}}, \bibinfo {author} {\bibfnamefont {K.}~\bibnamefont {Jiang}},
  \bibinfo {author} {\bibfnamefont {C.}~\bibnamefont {Scoggins}}, \bibinfo
  {author} {\bibfnamefont {D.}~\bibnamefont {Wingert}}, \bibinfo {author}
  {\bibfnamefont {D.}~\bibnamefont {Cubero}},\ and\ \bibinfo {author}
  {\bibfnamefont {S.}~\bibnamefont {Bali}},\ }\bibfield  {title} {\bibinfo
  {title} {Observation of stochastic resonance in directed propagation of cold
  atoms},\ }\href@noop {} {\bibfield  {journal} {\bibinfo  {journal} {Phys.
  Rev. Research}\ }\textbf {\bibinfo {volume} {4}},\ \bibinfo {pages} {043211}
  (\bibinfo {year} {2022})}\BibitemShut {NoStop}%
\bibitem [{\citenamefont {Metcalf}\ and\ \citenamefont {van~der
  Straten}()}]{metcalfbook}%
  \BibitemOpen
  \bibfield  {author} {\bibinfo {author} {\bibfnamefont {H.~J.}\ \bibnamefont
  {Metcalf}}\ and\ \bibinfo {author} {\bibfnamefont {P.}~\bibnamefont {van~der
  Straten}},\ }\href@noop {} {{\it {\bibinfo {title} {Cooling and
  Trapping}}}} (\bibinfo  {publisher}
  {Springer-Verlag},\ \bibinfo {address} {New York},\ \bibinfo
  {year} {1999})\BibitemShut {NoStop}
\bibitem [{\citenamefont {Hodapp}\ \emph {et~al.}(1995)\citenamefont {Hodapp},
  \citenamefont {Gerz}, \citenamefont {Furtlehner}, \citenamefont {Westbrook},
  \citenamefont {Phillips},\ and\ \citenamefont {Dalibard}}]{hodapp}%
  \BibitemOpen
  \bibfield  {author} {\bibinfo {author} {\bibfnamefont {T.~W.}\ \bibnamefont
  {Hodapp}}, \bibinfo {author} {\bibfnamefont {C.}~\bibnamefont {Gerz}},
  \bibinfo {author} {\bibfnamefont {C.}~\bibnamefont {Furtlehner}}, \bibinfo
  {author} {\bibfnamefont {C.~I.}\ \bibnamefont {Westbrook}}, \bibinfo {author}
  {\bibfnamefont {W.~D.}\ \bibnamefont {Phillips}},\ and\ \bibinfo {author}
  {\bibfnamefont {J.}~\bibnamefont {Dalibard}},\ }\bibfield  {title} {\bibinfo
  {title} {Three-dimensional spatial diffusion in optical molasses},\
  }\href@noop {} {\bibfield  {journal} {\bibinfo  {journal} {Appl. Phys. B}\
  }\textbf {\bibinfo {volume} {60}},\ \bibinfo {pages} {135} (\bibinfo {year}
  {1995})}\BibitemShut {NoStop}%
\bibitem [{\citenamefont {Cubero}(2023)}]{cubero22}%
  \BibitemOpen
  \bibfield  {author} {\bibinfo {author} {\bibfnamefont {D.}~\bibnamefont
  {Cubero}},\ }\bibfield  {title} {\bibinfo {title} {Brillouin propagation
  modes of cold atoms undergoing sisyphus cooling},\ }\href@noop {} {\bibfield
  {journal} {\bibinfo  {journal} {Phys. Rev. E}\ }\textbf {\bibinfo {volume}
  {107}},\ \bibinfo {pages} {034102} (\bibinfo {year} {2023})}\BibitemShut
  {NoStop}%
\bibitem [{\citenamefont {Lutz}\ and\ \citenamefont
  {Renzoni}(2013)}]{lutren13}%
  \BibitemOpen
  \bibfield  {author} {\bibinfo {author} {\bibfnamefont {E.}~\bibnamefont
  {Lutz}}\ and\ \bibinfo {author} {\bibfnamefont {F.}~\bibnamefont {Renzoni}},\
  }\bibfield  {title} {\bibinfo {title} {Beyond boltzmann–gibbs statistical
  mechanics in optical lattices},\ }\href@noop {} {\bibfield  {journal}
  {\bibinfo  {journal} {Nature Physics}\ }\textbf {\bibinfo {volume} {9}},\
  \bibinfo {pages} {615} (\bibinfo {year} {2013})}\BibitemShut {NoStop}%
\bibitem [{\citenamefont {Cubero}\ \emph {et~al.}(2014)\citenamefont {Cubero},
  \citenamefont {Casado-Pascual},\ and\ \citenamefont {Renzoni}}]{cubero2014}%
  \BibitemOpen
  \bibfield  {author} {\bibinfo {author} {\bibfnamefont {D.}~\bibnamefont
  {Cubero}}, \bibinfo {author} {\bibfnamefont {J.}~\bibnamefont
  {Casado-Pascual}},\ and\ \bibinfo {author} {\bibfnamefont {F.}~\bibnamefont
  {Renzoni}},\ }\bibfield  {title} {\bibinfo {title} {Irrationality and
  quasiperiodicity in driven nonlinear systems},\ }\href@noop {} {\bibfield
  {journal} {\bibinfo  {journal} {Phys. Rev. Lett.}\ }\textbf {\bibinfo
  {volume} {112}},\ \bibinfo {pages} {174102} (\bibinfo {year}
  {2014})}\BibitemShut {NoStop}%
\bibitem [{\citenamefont {Cubero}\ and\ \citenamefont
  {Renzoni}(2018)}]{cubero2018}%
  \BibitemOpen
  \bibfield  {author} {\bibinfo {author} {\bibfnamefont {D.}~\bibnamefont
  {Cubero}}\ and\ \bibinfo {author} {\bibfnamefont {F.}~\bibnamefont
  {Renzoni}},\ }\bibfield  {title} {\bibinfo {title} {Asymptotic theory of
  quasiperiodically driven quantum systems},\ }\href@noop {} {\bibfield
  {journal} {\bibinfo  {journal} {Phys. Rev. E}\ }\textbf {\bibinfo {volume}
  {97}},\ \bibinfo {pages} {062139} (\bibinfo {year} {2018})}\BibitemShut
  {NoStop}%
\bibitem [{\citenamefont {Cubero}\ \emph {et~al.}(2018)\citenamefont {Cubero},
  \citenamefont {Robb},\ and\ \citenamefont {Renzoni}}]{cub18}%
  \BibitemOpen
  \bibfield  {author} {\bibinfo {author} {\bibfnamefont {D.}~\bibnamefont
  {Cubero}}, \bibinfo {author} {\bibfnamefont {G.}~\bibnamefont {Robb}},\ and\
  \bibinfo {author} {\bibfnamefont {F.}~\bibnamefont {Renzoni}},\ }\bibfield
  {title} {\bibinfo {title} {Avoided crossing and sub-fourier-sensitivity in
  driven quantum systems},\ }\href@noop {} {\bibfield  {journal} {\bibinfo
  {journal} {Phys. Rev. Lett.}\ }\textbf {\bibinfo {volume} {121}},\ \bibinfo
  {pages} {213904} (\bibinfo {year} {2018})}\BibitemShut {NoStop}%
\bibitem [{\citenamefont {Petsas}\ \emph {et~al.}(1999)\citenamefont {Petsas},
  \citenamefont {Grynberg},\ and\ \citenamefont {Courtois}}]{petsas99}%
  \BibitemOpen
  \bibfield  {author} {\bibinfo {author} {\bibfnamefont {K.}~\bibnamefont
  {Petsas}}, \bibinfo {author} {\bibfnamefont {G.}~\bibnamefont {Grynberg}},\
  and\ \bibinfo {author} {\bibfnamefont {J.-Y.}\ \bibnamefont {Courtois}},\
  }\bibfield  {title} {\bibinfo {title} {Semiclassical monte carlo approaches
  for realistic atoms in optical lattices},\ }\href@noop {} {\bibfield
  {journal} {\bibinfo  {journal} {Eur. Phys. J. D}\ }\textbf {\bibinfo {volume}
  {6}},\ \bibinfo {pages} {29} (\bibinfo {year} {1999})}\BibitemShut {NoStop}%
\bibitem [{\citenamefont {Schiavoni}\ \emph
  {et~al.}(2002{\natexlab{b}})\citenamefont {Schiavoni}, \citenamefont
  {Carminati}, \citenamefont {Sanchez-Palencia}, \citenamefont {Renzoni},\ and\
  \citenamefont {Grynberg}}]{renzoni02b}%
  \BibitemOpen
  \bibfield  {author} {\bibinfo {author} {\bibfnamefont {M.}~\bibnamefont
  {Schiavoni}}, \bibinfo {author} {\bibfnamefont {F.-R.}\ \bibnamefont
  {Carminati}}, \bibinfo {author} {\bibfnamefont {L.}~\bibnamefont
  {Sanchez-Palencia}}, \bibinfo {author} {\bibfnamefont {F.}~\bibnamefont
  {Renzoni}},\ and\ \bibinfo {author} {\bibfnamefont {G.}~\bibnamefont
  {Grynberg}},\ }\bibfield  {title} {\bibinfo {title} {Stochastic resonance in
  periodic potentials: Realization in a dissipative optical lattice},\
  }\href@noop {} {\bibfield  {journal} {\bibinfo  {journal} {Europhys. Lett.}\
  }\textbf {\bibinfo {volume} {59}},\ \bibinfo {pages} {493} (\bibinfo {year}
  {2002}{\natexlab{b}})}\BibitemShut {NoStop}%
\bibitem [{Note1()}]{Note1}%
  \BibitemOpen
  \bibinfo {note} {This definition of the probe detuning $\delta _p$ corrects a
  sign error in Ref.~\cite {samir22}.}\BibitemShut {Stop}%
\bibitem [{\citenamefont {Casado-Pascual}\ \emph {et~al.}(2013)\citenamefont
  {Casado-Pascual}, \citenamefont {Cubero},\ and\ \citenamefont
  {Renzoni}}]{cubero2013}%
  \BibitemOpen
  \bibfield  {author} {\bibinfo {author} {\bibfnamefont {J.}~\bibnamefont
  {Casado-Pascual}}, \bibinfo {author} {\bibfnamefont {D.}~\bibnamefont
  {Cubero}},\ and\ \bibinfo {author} {\bibfnamefont {F.}~\bibnamefont
  {Renzoni}},\ }\bibfield  {title} {\bibinfo {title} {Universal asymptotic
  behavior in nonlinear systems driven by a two-frequency forcing},\
  }\href@noop {} {\bibfield  {journal} {\bibinfo  {journal} {Phys. Rev. E}\
  }\textbf {\bibinfo {volume} {88}},\ \bibinfo {pages} {062919} (\bibinfo
  {year} {2013})}\BibitemShut {NoStop}%
  \bibitem [{\citenamefont {Kloeden}}\ and\ \citenamefont
  {Platen}(1992)]{kloeden}%
  \BibitemOpen
  \bibfield  {author} {\bibinfo {author} {\bibfnamefont {P.}~\bibnamefont
  {Kloeden}}\ and\ \bibinfo {author} {\bibfnamefont {E.}~\bibnamefont
  {Platen}},\ }\href@noop {} {{\it {\bibinfo {title} {Numerical Solution of
  Stochastic Differential Equations}}}}\ (\bibinfo  {publisher} {Springer},\
  \bibinfo {address} {NY},\ \bibinfo {year} {1992})\BibitemShut {NoStop}%
\end{thebibliography}
%

\end{document}